\makeatletter
\IfFileExists{papers/Elsevier/cas-dc.cls}{%
  \def\input@path{{papers/Elsevier/}}%
}{}
\makeatother

\documentclass[a4paper,fleqn]{cas-dc}

\usepackage[numbers]{natbib}
\usepackage{svg}
\usepackage{tikz}  
\usepackage{makecell}  
\usepackage{booktabs}  
\usepackage{siunitx}  

\IfFileExists{papers/bib/references.bib}{%
  \newcommand{\figdir}{figures}%
  \newcommand{\tikzdir}{tikz}%
  \newcommand{\bibsource}{papers/bib/references}%
  \newcommand{\bstsource}{papers/Elsevier/cas-model2-names}%
  \graphicspath{{papers/Elsevier/}}%
}{%
  \newcommand{\figdir}{../../figures}%
  \newcommand{\tikzdir}{../../tikz}%
  \newcommand{\bibsource}{../bib/references}%
  \newcommand{\bstsource}{cas-model2-names}%
}

\newcommand{\vcenteredinclude}[2][]{%
  \raisebox{-.5\height}{\includegraphics[#1]{#2}}%
}

\newcommand{\errorbar}[2]{%
    \begin{tikzpicture}[baseline=-0.6ex]
        \pgfmathsetmacro{\ratio}{min(#1/#2,1)}
        \fill[black!10] (0,0) rectangle (2.2,0.18);
        \fill[black!55] (0,0) rectangle ({2.2*\ratio},0.18);
        \node[anchor=west,font=\scriptsize]
            at (2.3,0.09) {\num{#1}};
    \end{tikzpicture}%
}

\newcommand{\componentbar}[3]{%
    \begin{tikzpicture}[baseline=-0.6ex]
        \pgfmathsetmacro{\ratio}{min(#2/#3,1)}
        \node[anchor=east,font=\scriptsize]
            at (-0.08,0.09) {$#1$};
        \fill[black!10] (0,0) rectangle (1.6,0.16);
        \fill[black!55] (0,0) rectangle ({1.6*\ratio},0.16);
        \node[anchor=west,font=\scriptsize]
            at (1.68,0.08) {\num{#2}};
    \end{tikzpicture}%
}

\newcommand{\shadedrow}[6]{%
    \cellcolor{#1}#2 &
    \cellcolor{#1}#3 &
    \cellcolor{#1}#4 &
    \cellcolor{#1}#5 &
    \cellcolor{#1}#6%
}

\def\tsc#1{\csdef{#1}{\textsc{\lowercase{#1}}\xspace}}
\tsc{HBM}
\tsc{AFT}

\begin{document}
\let\WriteBookmarks\relax
\def\floatpagepagefraction{1}
\def\textpagefraction{.001}

\shorttitle{AFT Neural Function Approximators for 1D Nonlinear Force Laws}
\shortauthors{M. Goldack et~al.}

\title [mode = title]{AFT Neural Function Approximators for 1D Nonlinear Force Laws}

\author[1]{Miriam Goldack}[orcid=0009-0009-3624-721X]
\cormark[1]
\ead{m.goldack@tu-berlin.de}
\credit{Methodology, Software, Investigation, Visualization, Writing – original draft, Writing – review \& editing}

\author[2]{Johann Groß}[orcid=0000-0003-0289-1938]
\credit{Resources}

\author[2]{Malte Krack}[orcid=0000-0001-7154-1160]
\credit{Supervision, Funding acquisition, Writing – review \& editing}

\author[1]{Merten Stender}[orcid=0000-0002-0888-8206]
\credit{Supervision, Funding acquisition, Writing – review \& editing} 

\affiliation[1]{organization={Technische Universität Berlin, Chair of Cyber-Physical Systems in Mechanical Engineering},
    addressline={Straße des 17. Juni 135}, 
    postcode={10623},
    postcodesep={},
    city={Berlin},
    country={Germany}}

\affiliation[2]{organization={University of Stuttgart, Institute of Aircraft Propulsion Systems},
    addressline={Pfaffenwaldring 6}, 
    postcode={70569},
    postcodesep={}, 
    city={Stuttgart},
    country={Germany}}

\cortext[cor1]{Corresponding author}

\begin{abstract}
Nonlinear contacts and friction strongly influence the vibration response of assembled structures, but their accurate numerical treatment is computationally demanding. The harmonic balance method is widely used to compute periodic steady-state responses, yet the required alternating frequency–time scheme becomes costly for nonsmooth and hysteretic nonlinearities and must be repeated throughout the nonlinear solution process. Here we show that this procedure can be replaced by neural networks that directly map displacement Fourier coefficients to nonlinear force coefficients and provide the corresponding Jacobian through automatic differentiation. The surrounding solver and continuation algorithms remain unchanged for the computation of frequency response curves.

The neural networks exclusively learn individual nonlinear elements rather than complete system responses. Physics-based nondimensionalization and phase normalization facilitate the learning process and enable a single trained network to cover a wide range of parameter combinations.
Building on the cubic spring, unilateral spring, and Jenkins elements considered here, the approach points toward a reusable library of nonlinear-element surrogates that can be combined in arbitrary number and location within a mechanical system.
By bypassing the iterative force evaluation in time domain, the method offers favorable computational scaling for high-resolution analyses and systems with many nonlinear elements.
\end{abstract}



\maketitle

\section{Introduction}\label{Introduction}
Nonlinear effects play a crucial role in the periodic steady-state response of engineering structures, particularly in the vicinity of resonances. In assembled structures, mechanical joints with frictional interfaces contribute substantially to the effective stiffness and damping and may therefore strongly affect vibration amplitudes~\cite{Gaul1997, brake_mechanics_2018}. Their accurate representation is especially relevant for turbomachinery, where numerous contact interfaces, uncertain contact conditions, and a wide range of operating conditions must be considered during vibration prediction~\cite{krack_2017}. At the same time, high-fidelity contact models remain computationally demanding and are consequently often replaced by strongly simplified descriptions~\cite{brake_mechanics_2018}. 
One source of this computational complexity is the hysteretic behavior associated with dry friction. The hysteresis models are rate-independent, meaning that their force response depends on the direction of the relative motion, but not on the rate at which a given displacement path is traversed. The resulting history dependence and inherent non-smoothness pose particular challenges for numerical force evaluation.
Beyond dry friction at contact interfaces, hysteretic models are also used for inelastic material behavior and arise in other engineering contexts, such as history-dependent fluid flow attachment and separation. Fast and scalable nonlinear response analyses are therefore of interest across a broad range of engineering applications, enabling more realistic nonlinear models as well as computationally demanding tasks such as broad parameter studies, uncertainty quantification, and design optimization.

The Harmonic Balance Method (HBM) is an established approach for computing periodic steady-state responses of nonlinear systems and is widely applied in nonlinear structural dynamics, including high-fidelity models of frictionally coupled bladed disks~\cite{krack_harmonic_2019,krack_2017}. The solution and the nonlinear forces are represented by truncated Fourier series, converting the governing equations from time domain into a system of algebraic equations in the frequency domain. Combined with numerical path continuation, the HBM enables the efficient tracking of nonlinear frequency response curves, including multiple-valued and strongly curved solution branches.

A crucial step within the HBM is the determination of the Fourier coefficients of the nonlinear forces as a function of the Fourier coefficients of the displacement vector. Since nonlinear constitutive relations are formulated in the time domain, the forces are commonly computed using the alternating frequency--time (AFT) scheme~\cite{cameron:hal-01333697}. Through the inverse Fourier transform, the displacement is obtained on a discrete time grid, the nonlinear forces are evaluated, and the resulting force time signal is transformed back into the frequency domain. These operations are repeated during every nonlinear solver iteration and at every point on the solution curve. Nonsmooth and hysteretic nonlinearities, such as unilateral contact and dry friction, require a high temporal resolution to capture sharp transitions and limit aliasing errors. In addition, history-dependent force laws require sequential evaluation along the time grid within each nonlinear element. Although independent elements can be evaluated in parallel, the associated effort increases with the temporal resolution and is compounded in systems containing many nonlinear interfaces.

Several strategies have been proposed to reduce the computational effort of nonlinear frequency-response analysis. In~\cite{woiwode_2020}, the predictor--corrector AFT is compared with a fully frequency-domain asymptotic numerical method, highlighting the latter for sufficiently smooth nonlinearities while favoring AFT for nonsmooth contact problems. In~\cite{Gross2024}, the sequential nature of path continuation is addressed through a multi-fidelity strategy that enables independent, parallel correction of high-fidelity solution points. Parallel HBM implementations have also been developed for large-scale structural models, including domain-decomposition approaches in which subdomains are solved in parallel and coupled through interface conditions~\cite{Blahos.2020,Saponaro.2025,renkin2026}. These approaches target different parts of the solution process and are complementary to accelerating the nonlinear force evaluation itself.

Derivative evaluation within HBM has been addressed using automatic differentiation (AD). Early work employed AD for a discrete adjoint harmonic-balance solver in turbomachinery optimization~\cite{HUANG2014}, while more recent structural-dynamics implementations use forward AD with dual numbers~\cite{s_martins_python_2023} or reverse-mode AD in differentiable computing frameworks~\cite{chen_harmonic_2026}. AD-based Jacobians are also available in the open-source \texttt{pyHarm} framework~\cite{pyharm}. 
These approaches facilitate the consistent and automated differentiation of the harmonic-balance residual. However, the computational cost associated with the sequential time-domain evaluation of history-dependent nonlinear forces remains.

Machine learning has likewise been combined with HBM and nonlinear structural dynamics. Neural networks have been used to parametrize periodic solutions directly~\cite{ge_neural_2023}, to represent nonlinear restoring forces within harmonic-balance-based identification~\cite{liu_deep_2026}, and as frequency-domain device models embedded in HBM simulations~\cite{ramella_frequency-domain_2025}. Beyond HBM, structure-preserving neural differential operators and differentiable modal formulations have been proposed for nonlinear structural dynamics~\cite{Najera-Flores2023,Zheleznov_2026}. However, to the authors' knowledge, no previous work has directly replaced the AFT nonlinear-force evaluation in structural HBM by a neural frequency-domain surrogate.

In this work, the AFT scheme is replaced by a neural network that directly maps the Fourier coefficients of the displacement to the corresponding nonlinear force coefficients, while the Jacobian of the learned force mapping is obtained by AD. By removing explicit model parameter dependencies through nondimensionalization and exploiting the time invariance of nonlinear elements through phase normalization, the learning problem is reduced to a simpler and less redundant mapping. This facilitates training with comparatively simple network architectures and limited training data while enabling the resulting fixed networks to be applied across a wide range of parameter configurations.
Importantly, the neural networks are formulated at the level of individual nonlinear elements rather than at the level of the response of a particular mechanical system. They can therefore be integrated modularly into the existing HBM framework, leaving the residual equations, Newton-type iterations, and path-continuation procedure unchanged. This element-level formulation points toward a reusable library of nonlinear base-element surrogates that are trained once for a given harmonic truncation order and can subsequently be deployed without retraining across different parameter configurations, in arbitrary numbers, and at arbitrary locations within a mechanical system. For a fixed network architecture, the resulting evaluation does not depend on the number of AFT time samples and provides favorable scaling potential for high temporal resolutions and systems containing many nonlinear elements.

The proposed approach is investigated for a cubic spring, a unilateral spring, and a hysteretic Jenkins element, representing smooth, nonsmooth, and history-dependent nonlinear behavior, respectively. Its performance is assessed on the level of the predicted force coefficients and Jacobians as well as on the system level by comparing the resulting frequency response curves, Jacobian conditioning, and solver convergence behavior with AFT-based reference solutions. The results demonstrate that the computationally critical AFT evaluation can be replaced while preserving the relevant accuracy and numerical behavior of the established HBM solution procedure.

\section{Method}

\subsection{Harmonic Balance Method with Alternating Frequency-Time Scheme}

The HBM is a frequency-domain method for computing periodic steady-state solutions of nonlinear mechanical systems. In the present work, a general $d$-degree-of-freedom system containing $n_{\mathrm{nl}}$ local nonlinear elements is considered, 
\begin{equation}
    \mathbf{M} \ddot{\mathbf{q}}(t) + \mathbf{D}\dot{\mathbf{q}}(t) + \mathbf{K} \mathbf{q}(t) + \sum_{e=1}^{n_{\mathrm{nl}}} \mathbf{w}_e  f_{\text{nl},e}(t) = \mathbf{f}_{\text{ex}}(t) \quad.
\end{equation}
Here $\mathbf{M}$, $\mathbf{D}$, and $\mathbf{K} \in \mathbb{R}^{d\times d}$ denote the mass, linear damping, and stiffness matrices, respectively, $\mathbf{q} \in \mathbb{R}^{d}$ is the vector of generalized coordinates, and $\mathbf{f}_{\text{ex}} \in \mathbb{R}^{d}$ is an external excitation. The scalar force $f_{\text{nl}, e}$ of the $e$-th local nonlinear element is mapped to the global coordinates by the vector $\mathbf{w}_e \in \mathbb{R}^d$. 

For a harmonic excitation $\mathbf{f}_{\text{ex}}=\mathbf{f}_0\cos(\Omega t)$ with excitation frequency $\Omega$ and period $T=2\pi /\Omega$, the present work considers $T$-periodic steady-state responses. HBM is not restricted to responses that are synchronized with excitation. Suitable choices of one or multiple base frequencies also permit the representation of subharmonic and quasi-periodic responses~\cite{Hetzler_2023}, but these are beyond the scope of the present work.

The global displacement is approximated by a truncated Fourier series,
\begin{equation}
    \mathbf{q}(t) \approx \mathbf{q}_H(t) = \mathbf{a}_0 + \sum_{h=1}^H (\mathbf{a}_h\cos{(h\Omega t)} + \mathbf{b}_h\sin{(h\Omega t)}) \quad ,
\end{equation}
with truncation order $H$, where $\mathbf{a}_0$, $\mathbf{a}_h$, and $\mathbf{b}_h \in \mathbb{R}^d$ are real-valued Fourier coefficient vectors. For the $e$-th local nonlinear element, with relative displacement $q_e(t)=\mathbf{w}_e^\top \mathbf{q}(t)$, the corresponding local Fourier coefficients are collected in $\hat{\mathbf{q}}_{e,H} = [a_{e,0}, a_{e,1}, b_{e,1}, \dots, a_{e,H}, b_{e,H}]^{\top}$ $\in \mathbb{R}^{2H+1}$.

Inserting the truncated ansatz $\mathbf{q}_H(t)$ into the equation of motion does not, in general, satisfy the equation pointwise in time and leaves a residual
\begin{equation}
\begin{split}
    \mathbf{r}_H(t) = {}
    &\mathbf{M} \ddot{\mathbf{q}}_H(t) + \mathbf{D} \dot{\mathbf{q}}_H(t) + \mathbf{K} \mathbf{q}_H(t) \\
    &+ \sum_{e=1}^{n_{\mathrm{nl}}}\mathbf{w}_e f_{\mathrm{nl},e}(t) - \mathbf{f}_{\mathrm{ex}}(t) \quad .
\end{split}
\end{equation}

The HBM enforces the Fourier coefficients $\hat{\mathbf{r}}_H$ of this residual to vanish up to harmonic order $H$. Transforming linear terms $\mathbf{M} \ddot{\mathbf{q}}_H(t) + \mathbf{D} \dot{\mathbf{q}}_H(t) + \mathbf{K} \mathbf{q}_H(t)$ into the real cosine-sine representation in the frequency domain yields $\mathbf{S}(\Omega)\mathbf{\hat{q}}_H$ with the linear dynamic stiffness matrix $\mathbf{S}(\Omega) = \operatorname{diag} \bigl(\mathbf{S}_0, \mathbf{S}_1(\Omega), ..., \mathbf{S}_H(\Omega)\bigr)$. The static contribution is given by $\mathbf{S}_0 =\mathbf{K}$ and the block corresponding to the harmonics $h\geq 1$ reads
\begin{equation}
    \mathbf S_h(\Omega) = 
    \begin{bmatrix}
        \mathbf{K} - (h\Omega)^2 \mathbf{M} & h\Omega \mathbf{D} \\
        -h\Omega \mathbf{D} & \mathbf{K} - (h\Omega)^2 \mathbf{M}
    \end{bmatrix}
    \quad .
\end{equation}
The HBM equations therefore form a nonlinear algebraic system of $d(2H+1)$ equations for the $d(2H+1)$ unknown displacement Fourier coefficients
\begin{equation}
    \mathbf{\hat{r}}_H = \mathbf{S}(\Omega)\mathbf{\hat{q}}_H + \mathbf{\hat{f}}_{\mathrm{nl},H} - \mathbf{\hat{f}}_{\mathrm{ex},H} \overset{!}{\approx} \mathbf{0} \quad , \label{eq:frequency-domain_residual}
\end{equation}
which is solved iteratively using a numerical root-finding method, typically based on a Newton-type method. Here, $\mathbf{\hat{f}}_{\mathrm{nl},H} \in \mathbb{R}^{d(2H+1)}$ denotes the assembled Fourier coefficients of all local nonlinear force contributions.

In the general case, the nonlinear force coefficients of a local nonlinear element $\mathbf{\hat{f}}_{\mathrm{nl},e,H}$ in the frequency domain are not computable in closed form and must be obtained using an alternating frequency-time (AFT) scheme \cite{cameron:hal-01333697}. In the following, the three steps of the AFT scheme are described together with their asymptotic computational complexity. Big-$\mathcal{O}$ notation is used to denote an upper bound on the computational cost with respect to the number of retained harmonics $H$ and time samples $N$. 

First, the local displacement coefficients $\hat{\mathbf{q}}_{e,H}$ are transformed into $N$ discrete time samples of displacement $\mathbf{q}_{e,N}$ and velocity $\mathbf{\dot{q}}_{e,N}$ on the equidistant grid $t_j=j\Delta t$, $j=0,\dots,N-1$:
\begin{equation}
    \begin{split}
        \mathbf{q}_{e,N} = \mathbf{E}_{NH}\mathbf{\hat{q}}_{e,H} \quad &\Rightarrow \quad \mathcal{O}(NH) \quad ,\\
    \mathbf{\dot{q}}_{e,N} = \mathbf{E}_{NH} \left(\mathbf{\Lambda}_\Omega\mathbf{\hat{q}}_{e,H} \right) \quad &\Rightarrow \quad \mathcal{O}(NH+H) = \mathcal{O}(NH)\ .
    \end{split}
\end{equation}
Here, $\mathbf{E}_{NH}$ denotes the real Fourier synthesis matrix which is vectorizable and parallelizable over $N$, and $\mathbf{\Lambda}_{\Omega}$ is the spectral differentiation matrix mapping displacement coefficients to velocity coefficients which is vectorizable and parallelizable over $H$. 

Second, the nonlinear force is evaluated at the discrete time instances. The form and computational cost of this evaluation depend on whether the nonlinear force law is memoryless or history-dependent.
For memoryless nonlinearities with
\begin{equation}
    \mathbf{f}_{\mathrm{nl}, e, N} = f_{\mathrm{nl}}(\mathbf{q}_{e,N}, \dot{\mathbf{q}}_{e,N}) \quad \Rightarrow \quad \mathcal{O}(N) \quad ,
\end{equation}
the force at each time instance depends only on the instantaneous displacement and velocity. The $N$ evaluations are therefore independent and can be vectorized or evaluated in parallel. The required number of time samples, however, is closely related to the number of relevant harmonics. Smooth quantities can generally be represented accurately with fewer harmonics. In particular, for polynomial nonlinearities of degree $p$, the highest generated harmonic is $pH$. Hence, a finite number of time samples can be selected to avoid aliasing. For the cubic nonlinearity considered here, $N \geq 4H+1$ is sufficient.

Nonsmooth nonlinearities with sharp transitions such as contact onset generally do not have a finite highest harmonic. Consequently, a finite temporal discretization cannot represent the complete force spectrum, and $N$ must be increased to capture the relevant time-domain features and reduce aliasing. In the present nonsmooth application cases, temporal resolutions on the order of $N=2^{10}$ are employed.

Hysteresis models additionally require to account for the evolution of internal state variables. Rather than introducing these states as additional Fourier unknowns, they are treated implicitly by a marching procedure in time domain. Introducing an internal state variable $\mathbf{z}_e$ and constitutive evolution law $\psi$, the nonlinear force and the evolution of the internal state can generally be expressed as
\begin{equation}
    \mathbf{f}_{\mathrm{nl},e} = f_{\mathrm{nl}} \left( \mathbf{q}_e, \dot{\mathbf{q}}_e, \mathbf{z}_e \right), \qquad \dot{\mathbf{z}}_e = \psi \left( \mathbf{z}_e, \mathbf{q}_e, \dot{\mathbf{q}}_e \right) \quad .
\end{equation}
Starting from an initial state, the internal state and nonlinear force are advanced sequentially along the reconstructed displacement history on the discrete time grid. Consequently, the evaluation cannot generally be parallelized over the time samples, although independent nonlinear elements may still be evaluated in parallel. In addition, a suitable initialization is required to establish a periodic steady state of the internal variables. If $P$ periods are traversed for this purpose, the force-evaluation cost scales as $\mathcal{O}(PN)$. In the application case considered here, two periods are sufficient.

\begin{figure*}[t]
    \centering
    \includegraphics[width=\textwidth]{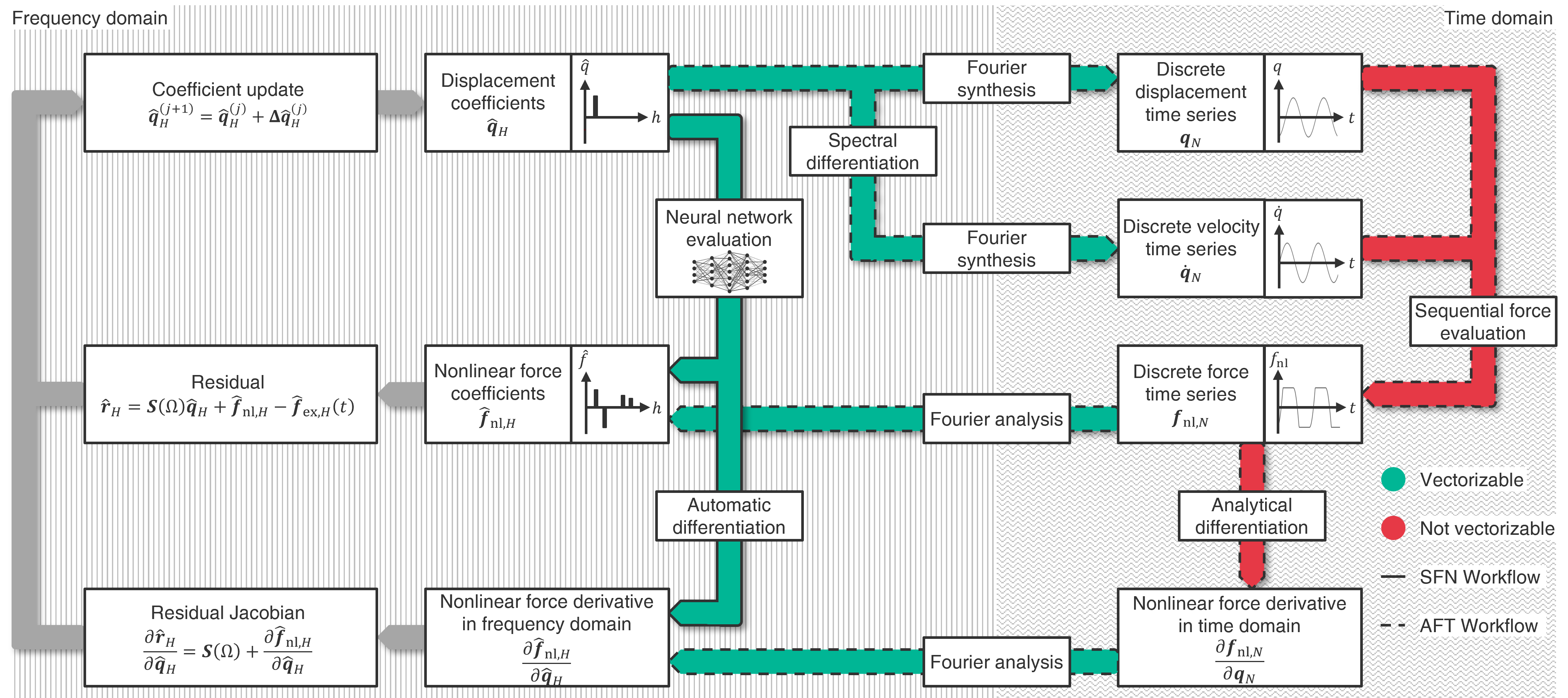}
    \caption{Comparison of the AFT- and SFN-based workflows within the HBM. Green and red indicate vectorizable and non-vectorizable operations, respectively; solid and dashed borders denote the SFN and AFT workflows.}
    \label{fig:hbm_overview}
\end{figure*}

The third AFT step comprises projecting the resulting force time series $\mathbf{f}_{\mathrm{nl},e,N}$ onto the retained Fourier basis:
\begin{align}
    \mathbf{\hat{f}}_{\mathrm{nl},e,H}^{\mathrm{AFT}} = E_{HN} \mathbf{f}_{\mathrm{nl},e,N} \quad &\Rightarrow \quad \mathcal{O}(NH) \quad ,
\end{align}
where $\mathbf{E}_{HN}$ denotes the real Fourier analysis matrix. Instead of explicit matrix multiplication with $\mathcal{O}(NH)$, transformations between frequency and time domains can be evaluated using FFT algorithms with a computational complexity of $\mathcal{O}(N\log N)$, whose butterfly operations are parallelizable and vectorizable within each of the $\mathrm{log}N$ sequential FFT stages.

For compactness, the computation of the nonlinear force through the complete AFT scheme and its combined work estimate $W_{\hat{\mathbf{f}},\,\mathrm{AFT}}$, considering a sequential nonlinear force evaluation in time domain over $P$ periods, are denoted by 
\begin{equation}
\begin{split}
    \mathbf{\hat{f}}_{\mathrm{nl},e,H}^{\mathrm{AFT}} = \text{AFT}(\mathbf{\hat{q}}_{e,H}) \quad \Rightarrow \quad &W_{\hat{\mathbf{f}},\,\mathrm{AFT}} \propto 3NH + H + PN \\
    & \in\, \mathcal{O}(NH + PN) \quad . 
\end{split}
\end{equation}
Here, the individual terms retain the contributions of the dominant computational steps rather than representing an exact FLOP count. Correspondingly, the work estimate of an FFT-based AFT scheme is $W_{\hat{\mathbf{f}},\,\mathrm{AFT}} \propto 3N \log N + H + PN \in \mathcal{O}(N\log N + H + PN)$.

Equation~\eqref{eq:frequency-domain_residual} forms a nonlinear implicit equation for the unknown displacement coefficient vectors $\mathbf{a}_0$, $\mathbf{a}_h$, and $\mathbf{b}_h$, collected in $\mathbf{\hat{q}}_H$. Its numerical solution commonly relies on Newton-type methods based on a local linearization of the residual. At an iterate $\mathbf{\hat{q}}_H^{(j)}$, this linearization reads
\begin{equation}
\hat{\mathbf r}_H
\left(\hat{\mathbf q}_H^{(j)}+\Delta\hat{\mathbf q}_H^{(j)}\right)
\approx
\hat{\mathbf r}_H\left(\hat{\mathbf q}_H^{(j)}\right)
+
\frac{\partial\hat{\mathbf r}_H}{\partial\hat{\mathbf q}_H}
\Bigg|_{\hat{\mathbf q}_H^{(j)}}
\Delta\hat{\mathbf q}_H^{(j)} \quad .
\end{equation}
Consequently, the residual Jacobian
\begin{equation}
    \frac{\partial \mathbf{\hat{r}}_H}{\partial \mathbf{\hat{q}}_H} = \mathbf{S}(\Omega) + \frac{\partial \mathbf{\hat{f}}_{\text{nl},H}}{\partial \mathbf{\hat{q}}_H} \quad 
\end{equation}
is a central quantity for the nonlinear solution procedure.

The required number of iterations depends on the initial guess and the convergence tolerance. In the considered examples, typically $n_{\mathrm{it}} \in \left\{1,\dots, 5\right\}$ Newton-type iterations were required per solution point, using the previously converged solution as an initial guess within the continuation procedure.

While $\mathbf{S}(\Omega)$ is directly available, the nonlinear contribution to the residual Jacobian requires differentiation of the nonlinear force mapping of each nonlinear element. Depending on the nonlinear force law and implementation, this derivative may be obtained analytically, semi-analytically, or by AD. The associated computational cost is therefore implementation-dependent. In the following, the estimate refers specifically to the semi-analytical AFT Jacobian used for the reference computations of the hysteresis model, where sensitivities are propagated in the time domain for each retained displacement coefficient. 

For a scalar nonlinear element, the $2H+1$ sensitivity directions are propagated over $PN$ time instances, resulting in a computational cost of $\mathcal{O}(PNH)$. The resulting time-domain sensitivity matrix is subsequently transformed back to the retained Fourier coefficients. Using the explicit Fourier analysis matrix, this operation scales as $\mathcal{O}(NH^2)$, whereas an FFT-based implementation scales as $\mathcal{O}(HN\log N)$.
Consequently, for the explicit transformation used in the present complexity estimate, 
\begin{equation}
\begin{split}
    & W_{\partial \hat{\mathbf{f}},\,\mathrm{AFT}} \in \, \mathcal{O}\left( P N H + NH^2\right) \quad .
\end{split}
\end{equation}
Each Newton-type iteration additionally requires the solution of a linearized system of size $d (2H+1)$. For a dense direct solver, the corresponding computational cost scales as $\mathcal{O}(d^3H^3)$. The work required to obtain a solution at a fixed excitation frequency with $n_{\mathrm{it}}$ Newton iterations can therefore be estimated as
\begin{equation}
\begin{split}
    W_{\mathrm{HBM}} \in \mathcal{O}\left( n_{\mathrm{it}} \left[ n_{\mathrm{nl}} \left( PNH + NH^2 \right) + d^3H^3\right] \right) \quad . 
\end{split}
\end{equation}
For a complete frequency response curve (FRC) computation with $n_\Omega$ points and a mean iteration count $\bar n_{\mathrm{it}}$, this becomes
\begin{equation}
\begin{split}
    W_{\mathrm{FRC}} \in \mathcal{O} \left( n_\Omega \bar n_{\mathrm{it}} \left[n_{\mathrm{nl}} \left(PNH + NH^2\right) + d^3H^3\right] \right) .
\end{split}
\end{equation}
The present work specifically targets the AFT-related per-iteration computational cost and, in particular, its dependence on the temporal resolution $N$, rather than $n_\Omega$ and $\bar n_{\mathrm{it}}$. The asymptotic scaling alone does not determine which contribution dominates at practically relevant values of $H$ and $N$. This implementation-dependent behavior is therefore examined separately in the runtime study in Section~\ref{sec:performance_comparison}.

\subsection{Spectral Force Network}
The repeated transformations between the frequency and time domains, the potentially expensive time-domain evaluation of nonlinear forces, and the numerical estimation of the Jacobian make the AFT scheme computationally demanding. To address these costs, we propose replacing the intermediate time-domain AFT evaluation by a learned frequency-domain mapping, implemented as a neural network hereafter referred to as the Spectral Force Network (SFN) $f_{e,\theta}:~\mathbb{R}^{2H+1}~\rightarrow~\mathbb{R}^{2H+1}$. The SFN approximates the mapping of the displacement coefficients $\hat{\mathbf{q}}_{e,H}=[a_0, a_1, b_1, \ldots, a_H, b_H]^\top$ to the nonlinear force coefficients $\hat{\mathbf{f}}_{\text{nl},e,H}=[A_0, A_1, B_1, \ldots, A_H, B_H]^\top$, while its Jacobian can be readily computed using automatic differentiation (AD).

Neural networks represent universal function approximators~\cite{Hornik1989} and are realized as computational graphs that express deeply nested functions as compositions of elementary operations. This structure decomposes complex mappings of inputs $x$ to outputs $\hat{y}$ into simple, localized operations with known partial derivatives. AD~\cite{rall1981} exploits this graph structure to propagate derivative information through the neural network using the chain rule. In forward mode, derivative information is propagated from the inputs to the outputs, whereas reverse mode propagates sensitivities from the outputs back to the inputs and forms the basis of backpropagation. Compared with symbolic differentiation, which can become cumbersome for high-dimensional problems, and numerical differentiation, which is affected by step-size-dependent approximation and round-off errors, AD provides an efficient means of computing derivatives of the neural network outputs with respect to its inputs up to machine precision.

The evaluation of the trained SFN, referred to as inference, is given by
\begin{equation}
    \hat{\mathbf{f}}_{\text{nl},e,H}^{\text{SFN}} = f_{\theta}(\hat{\mathbf{q}}_{e,H}) \approx \mathrm{AFT}(\hat{\mathbf{q}}_{e,H}) \quad ,
\end{equation}
where $f_{\theta}$ denotes the SFN with trainable parameters $\mathbf{\theta}$. In the proposed approach, one SFN is trained for one specific nonlinearity and a fixed HBM truncation order $H$, while the conceptual approach itself is not restricted to a particular nonlinearity. Typically, analyses in numerical structural dynamics consider one nonlinear force law or contact formulation at a time, rendering this specialization practical. The computational work of inference depends on the network architecture and size, as well as on the employed hardware and software implementation. Consequently, different SFNs may exhibit different absolute inference costs. However, for a fixed network architecture and implementation, this work is independent of the AFT time resolution $N$, since neither time-domain reconstructions nor time marching is required. The asymptotic dependence of the inference work on $N$ is therefore $W_{\hat{\mathbf{f}},\,\mathrm{SFN}} \in \mathcal{O}(1)$. Nonlinearities requiring a higher temporal resolution in the AFT may nevertheless be more difficult to approximate and could therefore require a larger neural network, increasing the absolute inference cost without changing its $\mathcal{O}(1)$ dependence on $N$.

Since the computation graph structure of the SFN is fully differentiable, the Jacobian contribution required in the Newton-type scheme is obtained directly by AD as
\begin{equation}
    \mathbf J_{\mathrm{nl},e,H}^{\mathrm{SFN}}
=\frac{\partial \hat{\mathbf{f}}_{\text{nl},e,H}^{\text{SFN}}}{\partial \hat{\mathbf{q}}_{e,H}} = \frac{\partial f_{\theta}}{\partial \hat{\mathbf{q}}_{e,H}} \approx \frac{\partial \hat{\mathbf{f}}_{\text{nl},e,H}}{\partial \hat{\mathbf{q}}_{e,H}} \quad ,
\end{equation}
without requiring force-law-specific derivative implementations.
The resulting Jacobian is analytically consistent with the function represented by the SFN and does not introduce approximation errors. However, its agreement with the Jacobian of the true nonlinear force mapping depends on how accurately the SFN captures not only the force coefficients but also their local variation with respect to the displacement coefficients.
The computational work of obtaining the full Jacobian via AD depends on the differentiation mode. Forward-mode AD constructs the Jacobian column-wise, whereas reverse-mode AD constructs it row-wise. Consequently, their relative efficiency depends primarily on the input and output dimensions. Since the SFN input and output dimensions are generally both $2H+1$, neither mode has an inherent dimensional advantage in the present setting. For reverse-mode AD, \cite{griewank2008} states that the gradient of a scalar-valued function can be evaluated at a computational cost of no more than approximately five times that of the corresponding function evaluation. Since each row of the SFN Jacobian corresponds to the gradient of one scalar output, a conservative upper bound for its complete Jacobian evaluation is therefore 
\begin{equation}
    \begin{split}
        W_{\partial \hat{\mathbf{f}},\mathrm{AD}}^{\mathrm{rev}} & \leq 5 (2H+1) W_{\hat{\mathbf{f}},\,\mathrm{SFN}} \\
        & \in \mathcal{O}(H) \quad .
    \end{split}
\end{equation}
This bound does not imply that the full Jacobian generally requires this exact amount of work, since intermediate computations may be reused and the actual cost depends on the AD implementation. For fixed $H$ and network architecture, however, $W_{\hat{\mathbf{f}},\,\mathrm{SFN}}$ and $W_{\partial \hat{\mathbf{f}},\mathrm{AD}}$ are both independent of the AFT time resolution $N$.
These properties make the approach particularly appealing for nonlinear contact forces that require high temporal resolution and whose Jacobians are not available analytically.

The SFN seamlessly replaces the AFT within the HBM, leaving the overall HBM solution procedure (including the continuation) unchanged while providing an efficient and differentiable evaluation of the nonlinear force at a constant and fixed computational cost.

\section{Test Systems}
Using the SFN promises benefits for systems with high temporal resolution requirements, such as nonsmooth systems. In order to verify the method, we first apply it to a mechanical oscillator with a cubic spring restoring force, namely the Duffing oscillator. As an example with nonsmooth transitions and a nonzero one-period force integral, we consider the unilateral spring. Lastly, as an example with hysteresis, we consider the Jenkins (or elastic dry friction) element, see Table~\ref{tab:application_infos}.

\begin{table*}[h]
\centering
\caption{Schematics of a single-mass damped oscillator with three different nonlinearities, force-displacement diagrams, and nonlinear force history for different numbers of time samples $N$.} 
\label{tab:application_infos}
\begin{tabular}{c c c}
\toprule
\hspace{1.5em} Cubic spring & \hspace{1.5em} Unilateral spring & Jenkins element \hspace{2.em} \\
\midrule
\hspace{2.2em} \includegraphics[width=4cm]{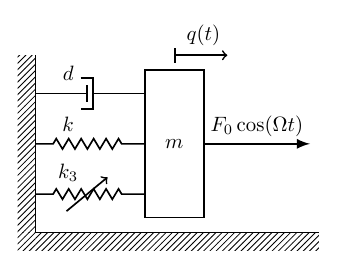} &
\hspace{1.6em} \includegraphics[width=4cm]{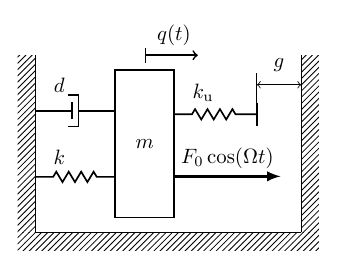} &
\includegraphics[width=4cm]{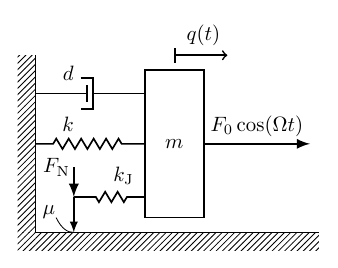} \hspace{2em} \\
\includegraphics[height=5.5cm]{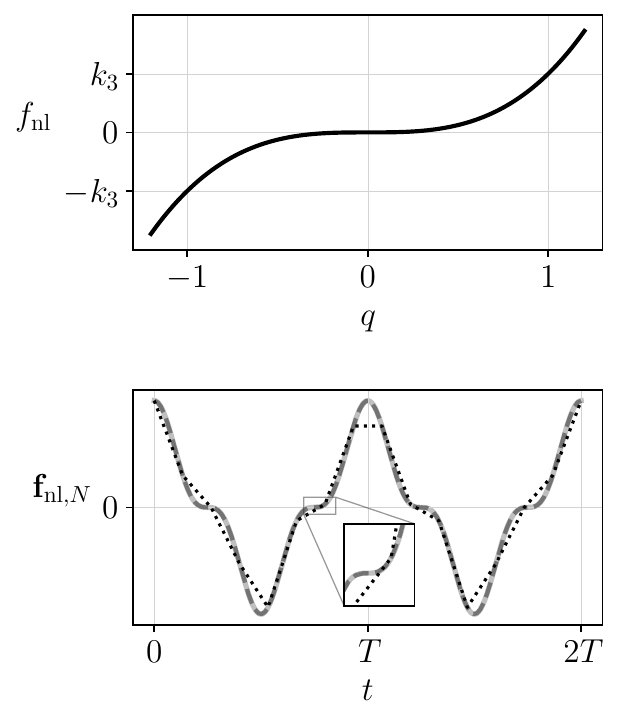} &
\includegraphics[height=5.5cm]{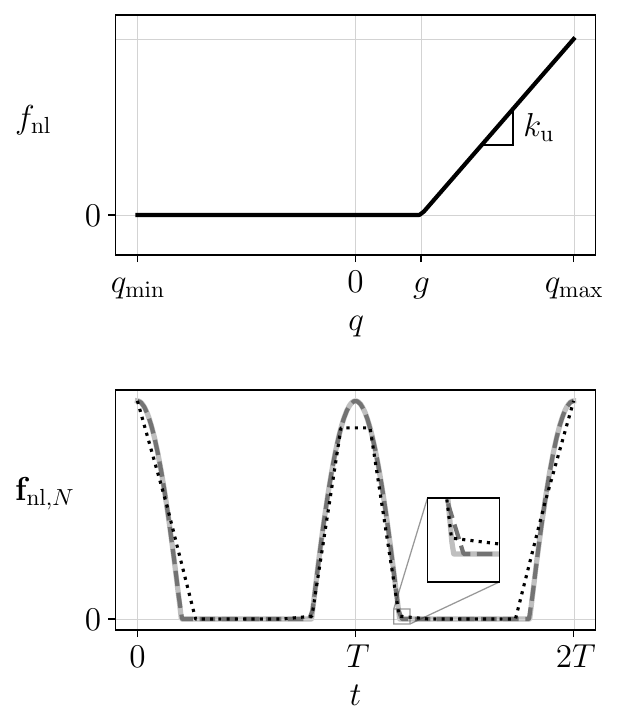} &
\includegraphics[height=5.5cm]{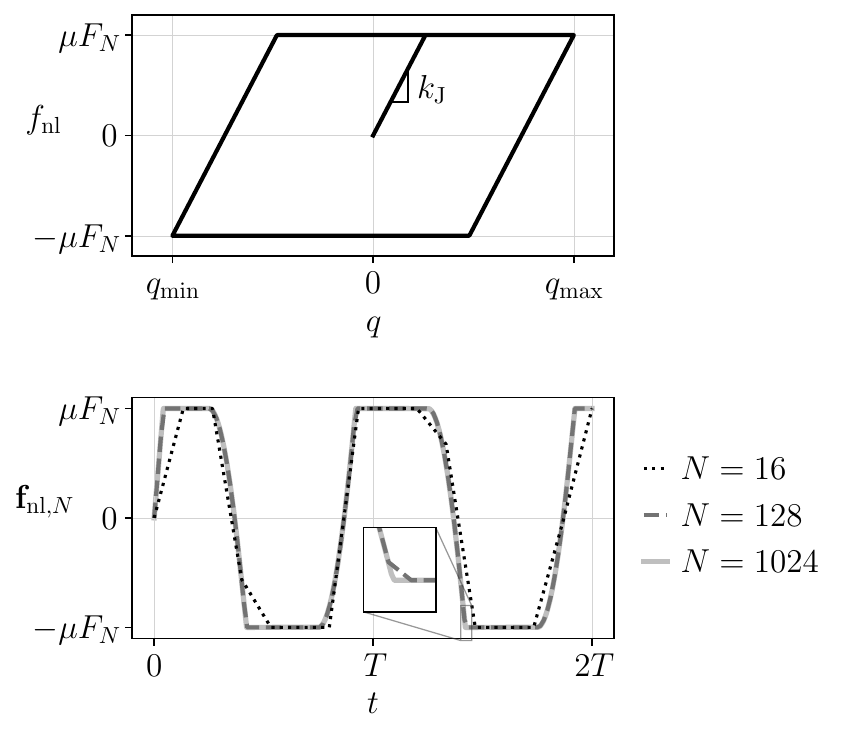} \\
\bottomrule
\end{tabular}
\end{table*}

\subsection{Cubic Spring}
The Duffing-type oscillator with a cubic spring nonlinearity, as illustrated in the left panel of Table\ref{tab:application_infos}, is a classical benchmark problem. Although the forced and damped system generally does not admit a simple closed-form solution, its qualitative dynamic behavior is well understood. This application case is particularly useful as a smooth reference problem, since the nonlinear force is differentiable. Positive and negative values of the cubic stiffness coefficient $k_3$ are considered, corresponding to hardening and softening behavior, respectively. The present analysis focuses on the primary resonance, with the Fourier expansion truncated at $H=3$ to retain the leading higher-harmonic contribution induced by the cubic nonlinearity while keeping the test case compact.

The cubic spring force law $f_{\mathrm{nl}} = k_3 q(t)^3$ has a complexity of $\mathcal{O}(N)$ for an $N$-element time-domain vector and can be evaluated pointwise, i.e. fully parallelized and vectorized over the time samples. Consequently, this case is not intended to demonstrate an immediate computational speedup, but rather serves as a transparent smooth benchmark for validating the proposed approach.

Since the cubic stiffness coefficient $k_3$ enters the nonlinear force law only as a scalar factor, it can be separated from the parameter-independent cubic mapping. The SFN is therefore trained to map the displacement coefficients to the Fourier coefficients of $q^3$, while the resulting nonlinear force coefficients and Jacobian are subsequently scaled by $k_3$. Consequently, the same trained network can be applied to both hardening and softening systems with different values of $k_3$, provided that the displacement coefficients remain within the trained input domain. Details are provided in Appendix~\ref{app:duffing_parameter_separation}.

Owing to the time invariance of the nonlinear force law, a phase shift of the displacement results in the corresponding harmonic-wise phase shift of the nonlinear force. The first-harmonic phase can therefore be removed before inference and restored analytically afterward. This eliminates redundant phase information, prevents the network from having to learn the underlying rotational equivariance, and concentrates the training data on physically distinct waveform shapes. Details of the phase normalization and the transformation of the predicted force coefficients and Jacobians back to the original phase are provided in Appendix~\ref{app:phase_normalization}. The parameter scaling commutes with the harmonic-wise phase transformation and can therefore be applied independently before or after phase restoration to both the force coefficients and their Jacobian.

Due to the odd symmetry of the restoring force, $f_{\mathrm{nl}}(-q) = -f_{\mathrm{nl}}(q)$, and the purely first-harmonic, zero-mean cosine excitation, the steady-state response considered here exhibits half-wave symmetry, provided that no static preload or asymmetric contribution is present (as in the case of a symmetry breaking bifurcation). Hence, the static and even-harmonic coefficients vanish, and only odd harmonics contribute to the displacement and nonlinear force. For the present test case with harmonic order $H=3$, the SFN input and output can therefore be restricted to the nonzero coefficients of the first and third harmonics. 

Including all physics-based pre- and postprocessing steps, namely parameter scaling $(\cdot)^*$, phase shift $(\cdot)'$ and omission of vanishing harmonic coefficients, the SFN input is $[a_1^{\prime}, a_3^{\prime}, b_3^{\prime}]$ and the corresponding parameter-independent output is $[A_1^{*,\prime}, B_1^{*,\prime}, A_3^{*,\prime}, B_3^{*,\prime}]$.

The training data were generated by independently sampling the phase-normalized coefficients from uniform distributions. While \(a_1'\) is restricted to non-negative values by the phase normalization, broad symmetric ranges are used for the higher-harmonic coefficients to avoid imposing prior assumptions on the encountered response states. The exact sampling domains and dataset sizes are provided in Appendix~\ref{app:training_data}, and the SFN architecture and training setup in Appendix~\ref{app:NN_specifications}.

\subsection{Unilateral Spring}
Secondly, a one-sided contact with an initial gap, as illustrated in the middle panel of Table~\ref{tab:application_infos}, is considered and hereafter denoted as a unilateral spring. It provides a simple model of intermittent contact by capturing the transition between free motion and contact onset, and therefore serves as a representative benchmark for nonsmooth contact nonlinearities.

The corresponding force law $f_{\mathrm{nl}} = k_\mathrm{u}(q-g) \mathbf{1}_{q \geq g}$ has a complexity of $\mathcal{O}(N)$ for an $N$-element time-domain vector and can be evaluated pointwise, i.e. fully parallelized and vectorized over the time samples. Here, $k_u$ and $g$ denote the unilateral spring stiffness and the gap, respectively, and $\mathbf{1}_{q \geq g}$ denotes the indicator function of the closed-contact state.

For displacements below the gap, the unilateral spring is inactive and its tangent stiffness is zero. Once contact is established, the tangent stiffness is constant $k_\mathrm{u}$. The resulting piecewise-linear force law is continuous but not differentiable at the transition between open and closed contact, rendering the force law nonsmooth. Consequently, an increased temporal resolution is required in the AFT evaluation to accurately resolve the contact transitions. In addition, the nonsmooth force law generally produces more pronounced higher-harmonic contributions than a smooth nonlinearity, making its Fourier representation more demanding. As a proof of concept, the harmonic truncation order is limited to $H=2$ in this application case, although higher harmonic orders may be required for an accurate representation of the contact force.

To remove the explicit dependence on the unilateral stiffness $k_{\mathrm{u}}$ and gap $g$, the displacement and nonlinear force are nondimensionalized using $g$ and $k_{\mathrm{u}}g$, respectively. The SFN therefore learns a parameter-independent contact mapping that can be applied to arbitrary positive values of $k_{\mathrm{u}}$ and $g$, provided that the dimensionless displacement coefficients remain within the trained input domain. Details are shown in Appendix~\ref{app:unilateral_scaling}.

Although the unilateral force law is asymmetric, it is time invariant and therefore equivariant with respect to phase shifts. The first-harmonic phase can consequently be removed before inference and restored analytically afterward for both the predicted force coefficients and their Jacobian, as described in Appendix~\ref{app:phase_normalization}.


For $H=2$, the SFN inputs are the nondimensional and phase-normalized coefficients $[a_0^{*,\prime}, a_1^{*,\prime}, a_2^{*,\prime}, b_2^{*,\prime}]$, with $b_1^{*,\prime}=0$ by construction. Correspondingly, the network outputs are the coefficients $[A_0^{*,\prime}, A_1^{*,\prime}, B_1^{*,\prime}, A_2^{*,\prime}, B_2^{*,\prime}]$.

The training data were generated by independently sampling the phase-normalized and nondimensionalized coefficients from uniform distributions. While $a_1^{*,\prime}$ is restricted to non-negative values by the phase normalization, broad symmetric ranges are used for the static and higher-harmonic coefficients to avoid imposing application-specific assumptions on the encountered response states. The exact sampling domains and dataset sizes are provided in Appendix~\ref{app:training_data}, and the SFN architecture and training setup in Appendix~\ref{app:NN_specifications}. A more application-specific sampling strategy exploiting additional physical information is discussed in Appendix~\ref{app:unilateral_physics_sampling}.

\subsection{Jenkins Element}
As a third application case, a Jenkins element is considered as a representative model for hysteretic tangential contact behavior. It captures the transition between sticking and sliding and is therefore widely used as a minimal model of frictional energy dissipation in jointed structures. The numerical model consists of a clamped-free Euler--Bernoulli beam with a localized Jenkins element and a harmonic point force. For illustration, the right panel of Table~\ref{tab:application_infos} shows a simplified single-degree-of-freedom schematic of the local Jenkins contact.

The nonsmooth stick--slip transitions place specific demands on both the temporal resolution of the AFT evaluation and the harmonic truncation of the HBM. While the generalized displacement remains comparatively smooth, the internal slider state and the resulting friction force are only continuous, and their time derivatives may change discontinuously at stick--slip transitions. Consequently, the nonlinear force exhibits sharper temporal features than the displacement driving it. While the overall hysteresis loop can already be reproduced with a moderate number of time samples, these local transitions require a sufficiently fine time discretization. This is illustrated in the third row of Table~\ref{tab:application_infos}, where increasing $N$ improves the local resolution of the nonlinear force history. The sharp changes in the force history also introduce pronounced higher-harmonic contributions. The Fourier approximation is therefore truncated at $H=3$ for the present benchmark, providing a compact proof of concept while retaining harmonics up to third order.

The resulting nonlinear force is path-dependent. For small relative displacements, sticking occurs and the contact force changes elastically with the tangential spring deformation. Once the force reaches the friction limit $\pm \mu F_{\mathrm{N}}$, sliding occurs. In a time-discrete AFT evaluation, this behavior can be represented by
\begin{equation}
\begin{split}
    f_{\mathrm{nl},i} = \mathrm{clip}\left(f_{\mathrm{nl},i-1} + k_{\mathrm{J}}(q_i-q_{i-1}), -\mu F_{\mathrm{N}}, \mu F_{\mathrm{N}} \right), \\
    i=1,\dots,N \quad ,
\end{split}
\end{equation}
where $q_i$ denotes the relative displacement at the $i$-th time sample. Since each force value depends on the preceding stick--slip state, the evaluation must be performed sequentially along the time grid and cannot be parallelized or vectorized over the time samples. Its computational complexity is $\mathcal{O}(PN)$, since $P$ periods have to be traversed to establish a steady-state hysteresis. In the present implementation, two periods are evaluated.

The displacement and force are nondimensionalized by the characteristic ratio of elastic spring stiffness to friction limit force $\frac{k_{\mathrm{J}}}{\mu F_\mathrm{N}}$ and the friction limit force $\mu F_\mathrm{N}$, respectively. The SFN therefore learns a parameter-independent dimensionless mapping that can be applied to different positive values of $k_\mathrm{J}$ and $\mu F_\mathrm{N}$, provided that the dimensionless inputs remain within the trained domain. Details are provided in Appendix~\ref{app:jenkins_scaling}.

Although the Jenkins force is history-dependent, its constitutive law has no explicit time dependence. Once the periodic internal state has been established, shifting the time origin produces the corresponding shift of the complete displacement--force trajectory, such that the periodic force mapping remains phase-equivariant. The first-harmonic phase can therefore be removed before inference and restored afterward; this operation is independent of the scalar nondimensionalization described above. Details are provided in Appendix~\ref{app:phase_normalization}.

For the symmetric Jenkins law under zero-mean, purely first-harmonic excitation, the considered steady-state response exhibits half-wave symmetry. Consequently, the static and even-harmonic coefficients of both the displacement and nonlinear force vanish. With $H=3$, only the first and third harmonics are therefore retained, which further reduces the SFN input and output dimensions.

For $H=3$, the SFN inputs are the reduced phase-normalized and nondimensional coefficients $[a_1^{*,\prime}, a_3^{*,\prime}, b_3^{*,\prime}]$, with $b_1^{*,\prime}=0$ by construction. The corresponding output is $[A_1^{*,\prime}, B_1^{*,\prime}, A_3^{*,\prime}, B_3^{*,\prime}]$. The predicted force coefficients and corresponding Jacobian are subsequently transformed back to the original phase and physical units by applying the inverse harmonic rotation and the appropriate scaling.

The training data were generated by independently sampling the phase-normalized and nondimensionalized coefficients from uniform distributions. While $a_1^{*,\prime}$ is restricted to non-negative values by the phase normalization, broad symmetric ranges are used for the higher-harmonic coefficients to avoid imposing application-specific assumptions on the encountered response states. The exact sampling domains and dataset size are provided in Appendix~\ref{app:training_data}, and the SFN architecture and training setup in Appendix~\ref{app:NN_specifications}.

\section{Results and Discussion}
All SFNs were implemented and trained in Python 3.12 using PyTorch 2.8 and integrated into the academic Matlab tool for nonlinear vibration analysis NLvib~\cite{krack_harmonic_2019}. 

The comparison between the classical AFT, which is considered as the reference and ground truth, and the SFN is evaluated on four levels. First, the accuracy is verified directly on the coefficient and Jacobian level by value-by-value comparison. Second, the SFN is tested as embedded into the HBM solver by comparing the resulting frequency response curves and Newton-type convergence behavior against the reference. Third, variations in the systems' physical parameters are considered to demonstrate the generalization capability of the proposed approach. Finally, computational performance is assessed in a Python benchmark.

\begin{table*}[h]
\caption{Comparison of the force-coefficient and Jacobian predictions along the considered frequency-response continuation. All global error metrics are evaluated using the same $n_\Omega$ points on the solution curve and identical inputs for the AFT and SFN evaluations. \label{tab:results}}
\centering
\begin{tabular}{l c c c}
\toprule
& Cubic spring & Unilateral spring & Jenkins element \\
\midrule
\addlinespace[0.8em]
\multicolumn{4}{@{}c}{Force coefficients: global relative $L^2$-error $\varepsilon_{\hat{\mathbf f},2}^{\mathrm{glob}}$} \\
\addlinespace[0.5em]
&
\errorbar{0.00135}{0.13635}
&
\errorbar{0.13635}{0.13635}
&
\errorbar{0.00314}{0.13635} 
\\
\addlinespace[0.8em]
\multicolumn{4}{@{}c}{Force coefficients: component-wise normalized RMSE} 
\\ 
\addlinespace[0.5em]
&
\makecell[l]{
    \componentbar{A_1}{0.00134}{1.68496}\\ 
    \componentbar{B_1}{0.88292}{1.68496}\\ 
    \componentbar{A_3}{0.00431}{1.68496}\\ 
    \componentbar{B_3}{1.68496}{1.68496}
} 
& 
\makecell[l]{
    \componentbar{A_0}{0.25498}{1.68496}\\ 
    \componentbar{A_1}{0.23456}{1.68496}\\ 
    \componentbar{B_1}{0.65452}{1.68496}\\ 
    \componentbar{A_2}{0.27431}{1.68496}\\ 
    \componentbar{B_2}{0.30011}{1.68496}
} 
& 
\makecell[l]{
    \componentbar{A_1}{0.00292}{1.68496}\\ 
    \componentbar{B_1}{0.00458}{1.68496}\\ 
    \componentbar{A_3}{0.03147}{1.68496}\\ 
    \componentbar{B_3}{0.01529}{1.68496}
} \\
\addlinespace[0.8em]
\multicolumn{4}{@{}c}{Jacobian: mean pointwise relative Frobenius norm error $\bar \varepsilon_{\mathbf J,F}$} \\
\addlinespace[0.5em]
&
\errorbar{0.00058}{0.16013} 
& 
\errorbar{0.16013}{0.16013} 
& 
\errorbar{0.04257}{0.16013} \\
\midrule
\addlinespace[0.8em]
\multicolumn{4}{@{}c}{Frequency response} \\
\addlinespace[0.5em]
&
\hspace{0.5em} \vcenteredinclude[height=3.cm]{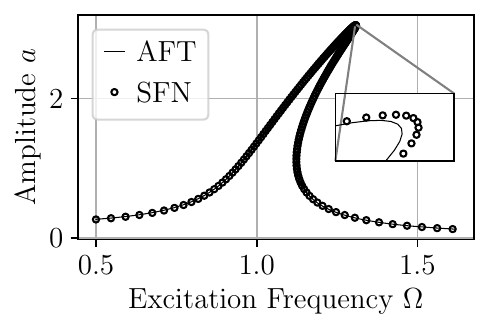} 
& 
\hspace{1em} \vcenteredinclude[height=3cm]{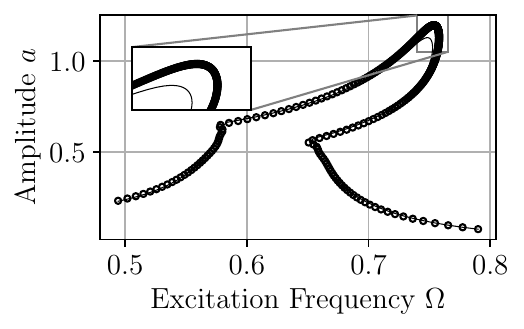} 
&
\makecell[l]{
    \hspace{0.5em} \vcenteredinclude[height=3.3cm]{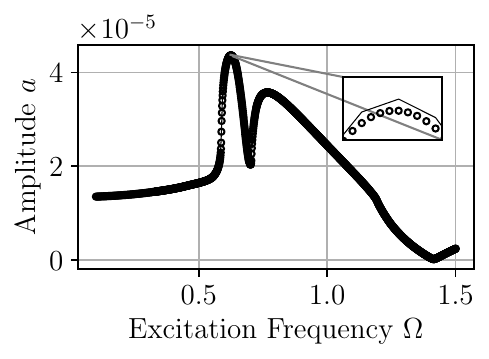} \\
    \vspace{-0.8em}
} \\
\addlinespace[0.8em]
\multicolumn{4}{@{}c}{Jacobian condition number} \\
\addlinespace[0.5em]
& 
\vcenteredinclude[height=3cm]{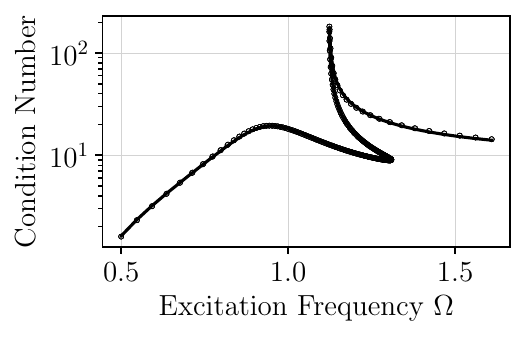} 
& 
\vcenteredinclude[height=3cm]{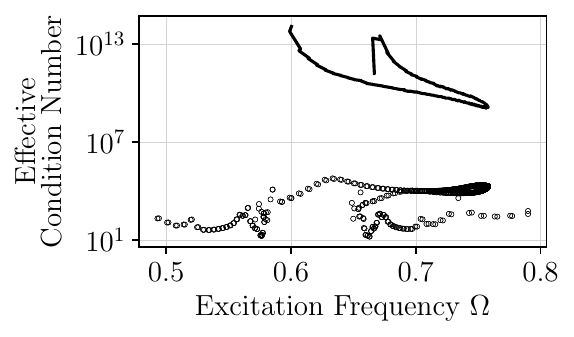} 
& 
\vcenteredinclude[height=3cm]{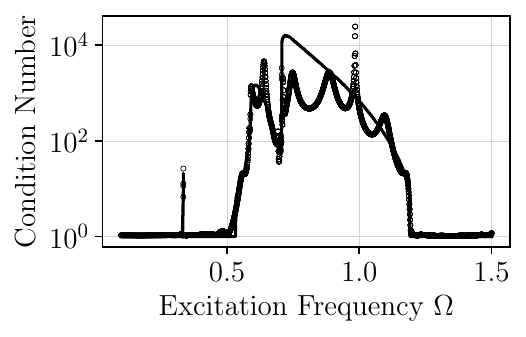} \\
\addlinespace[0.8em]
\multicolumn{4}{@{}c}{Newton iterations} \\
\addlinespace[0.5em]
&
\vcenteredinclude[height=6.5cm]{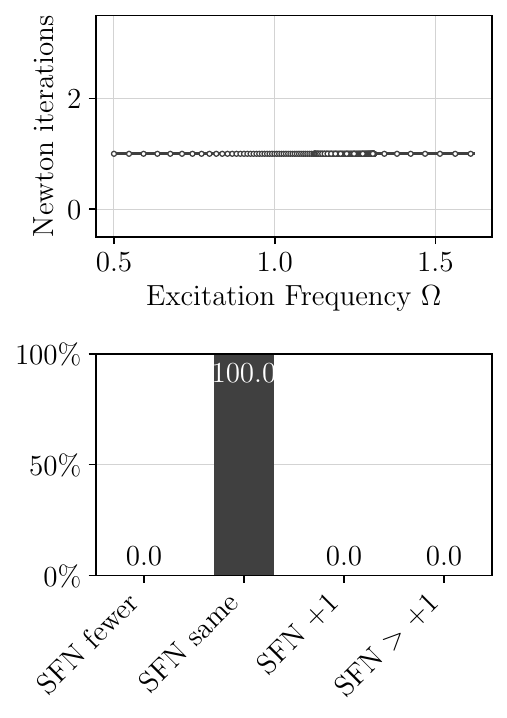} 
&  
\hspace{1em} \vcenteredinclude[height=6.5cm]{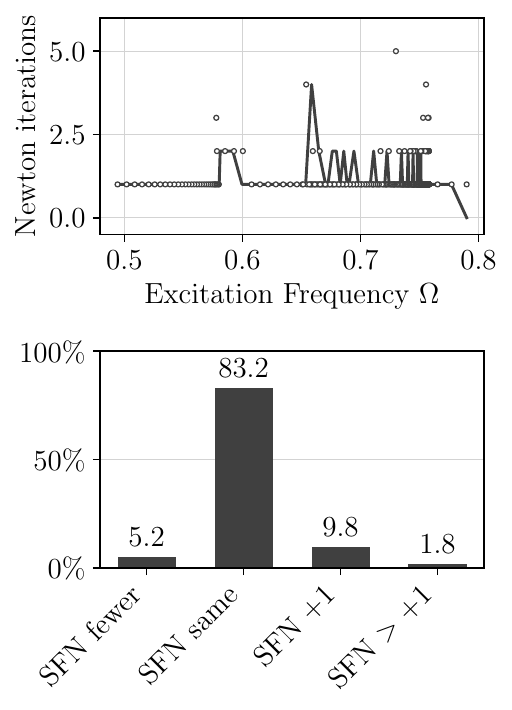} 
& 
\vcenteredinclude[height=6.6cm]{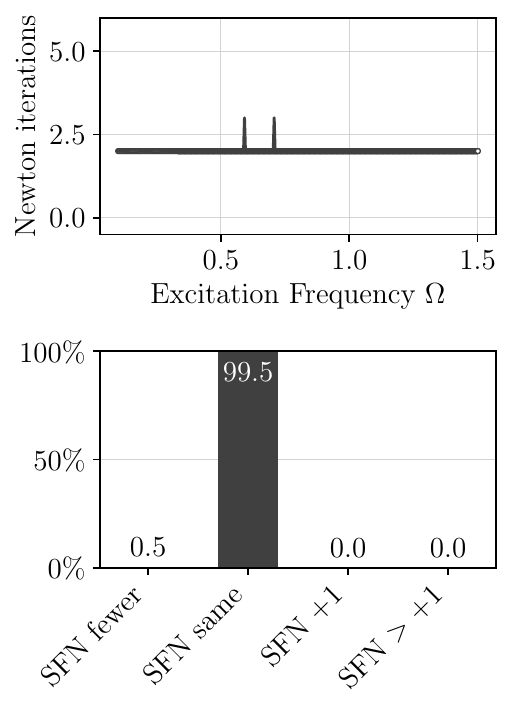} \\
\end{tabular}
\end{table*}

\subsection{Coefficient- and Jacobian-Level Accuracy}
The accuracy of the force-coefficient and Jacobian predictions is evaluated along a selected frequency-response continuation whose input points were not specifically used as training samples. All error metrics are evaluated using the same $n_\Omega$ points on the solution curve, with identical inputs for the AFT and SFN evaluations and a fixed continuation step size. The metrics reported in Table~\ref{tab:results} provide aggregate measures over the complete continuation path. Additionally, the corresponding frequency-resolved error measures are provided in Appendix~\ref{app:error_metrics}. Although the frequency-resolved errors exhibit local variations and isolated peaks, no systematic increase in error or persistent loss of accuracy along the considered frequency responses is observed.

First, the accuracy of the nonlinear force coefficients over the complete frequency response is quantified by the global relative $L^2$-error $\varepsilon_{\hat{\mathbf{f}},2}^{\mathrm{glob}}$, defined in Equation~(\ref{eq:rel_l2}) in Appendix~\ref{app:error_metrics}. It relates the $L^2$-norm of the coefficient errors accumulated over all $n_\Omega$ points on the solution curve to the corresponding norm of the AFT reference coefficients. The smallest error is observed for the cubic spring, while the unilateral spring exhibits a markedly larger error than the other two cases. This reflects the increased difficulty of approximating the nonsmooth unilateral contact mapping over the general training domain, particularly in regions with sharp changes in contact state or pronounced influence of higher harmonics.

While the global relative $L^2$-error $\varepsilon_{\hat{\mathbf{f}},2}^{\mathrm{glob}}$ assesses the overall accuracy of the predicted force-coefficient vectors, it may conceal component-specific deviations. Therefore, the component-wise normalized RMSE defined in Equation~(\ref{eq:nrmse}), Appendix~\ref{app:error_metrics}, is additionally evaluated for the individual harmonic coefficients. The normalized RMSE expresses the prediction error of each coefficient relative to its characteristic variation along the reference continuation, as quantified by its standard deviation. Consequently, coefficients that remain close to zero or vary only weakly can exhibit large normalized errors even when their contribution to the overall force error is limited. For the cubic spring, this is particularly apparent for the sine coefficients $B_1$ and $B_3$, with normalized RMSE values of $0.883$ and $1.685$, respectively, whereas the cosine coefficients exhibit substantially smaller errors. For the unilateral spring, the errors are more evenly distributed among the coefficients, ranging from approximately $0.23$ to $0.65$, with the largest value occurring for $B_1$. For the Jenkins element, all component-wise errors remain below $0.032$, with the largest value occurring for the third-harmonic coefficient $A_3$. These normalized component-wise measures are therefore interpreted together with the global force error and the resulting frequency-response agreement, rather than as standalone indicators of the relevance of individual coefficient errors.

The Jacobians obtained by AD are analytically consistent with the force-coefficient mapping represented by the SFN. However, discrepancies between the predicted and reference force mappings, particularly in their local variation with respect to the displacement coefficients, may lead to differences between the SFN-based and reference Jacobians. In addition, the use of smooth activation functions renders the SFN mapping continuously differentiable, such that discontinuous changes in the reference Jacobian associated with nonsmooth force laws are necessarily represented in a smoothed form. To assess this Jacobian-level accuracy, the SFN-based Jacobians are compared with the corresponding reference Jacobians along the selected frequency-response continuation. The error is quantified by the mean pointwise relative Frobenius norm error $\bar{\varepsilon}_{\mathbf{J},F}$, defined in Equation~(\ref{eq:frob_norm}), Appendix~\ref{app:error_metrics}. The resulting errors are $0.058\%$, $16.013\%$, and $4.257\%$ for the cubic spring, unilateral spring, and Jenkins element, respectively. The cubic spring shows very close agreement with the available analytical reference Jacobian, whereas the larger error for the unilateral spring reflects both the difficulty of reproducing the local derivatives of the nonsmooth contact mapping and the inherent smoothing introduced by the differentiable SFN representation. The Jenkins element exhibits an intermediate Jacobian error despite its history-dependent force law.

It should be noted, however, that Jacobian accuracy and suitability for the Newton solver are related but not equivalent: an accurate SFN-Jacobian may still be ill-conditioned, while a well-conditioned Jacobian is not necessarily accurate. Therefore, the conditioning of the resulting Newton systems is additionally examined in the subsequent solver-level validation.

\subsection{Solver-Level Assessment}
After verifying the SFN predictions themselves, the SFNs are integrated into the NLvib HBM solver suite, replacing exactly, and only, the AFT scheme. The PyTorch models are called from MATLAB through the Python interface. The same frequency-response cases and fixed continuation step sizes used in the preceding comparisons are retained for the solver-level assessment. The influence of replacing the AFT by the SFN is evaluated in terms of the frequency response curves, the Jacobian condition number and the Newton iterations required at each point on the solution curve, as summarized in Table~\ref{tab:results}. This assessment examines whether the learned nonlinear force mapping and its AD-based Jacobian are sufficiently accurate and numerically consistent to reproduce the convergence behavior of the original AFT-based formulation. 
The SFN itself represents a continuous and differentiable mapping of the displacement coefficients and no systematic deterioration of the SFN accuracy is observed along the continuation path. However, local discrepancies between the SFN and reference mappings may still affect the nonlinear solver differently at individual points on the solution curve. In particular, accurate force predictions do not by themselves guarantee equally accurate local derivatives at every solution point. 
The solver-level comparison therefore provides an additional test of whether such local discrepancies affect the Newton convergence when the SFN is embedded in the nonlinear solution procedure.

The resulting frequency response curves obtained with the SFN-based formulation closely reproduce the same solution branches, resonance locations, and overall response characteristics of the corresponding AFT-based HBM results for all three application cases over most of the considered frequency range. Small local deviations are observed in particularly sensitive regions (see zoomed views), such as the opening and closing of the unilateral contact or resonance peaks in general. For the unilateral spring, deviations occur near the contact transitions at $\Omega = 0.583$ and $0.654$ in Table~\ref{tab:results}, while a more pronounced discrepancy is observed around the resonance peak. The latter coincides with the largest local SFN prediction errors along the considered response, as shown in Figure~\ref{fig:error_metrics_vs_omega}. Near such regions, small differences in the predicted force coefficients or their local derivatives can lead to comparatively larger shifts in the converged response. These regions are also challenging for the AFT reference itself, since their accurate representation depends on the temporal resolution $N$ and the retained harmonic order $H$. Consequently, the observed FRC deviations reflect both the SFN approximation error and the local sensitivity of the nonlinear solution. As demonstrated in Appendix \ref{app:unilateral_physics_sampling}, the agreement can be substantially improved when application-specific physical information can be incorporated into the training-data sampling.

In addition to the frequency response curves, the conditioning of the Jacobians is analyzed along the frequency sweeps. Since each Newton step requires solving a linear system based on the current Jacobian, its condition number provides a diagnostic measure for the numerical robustness of the linearized solve, with smaller condition numbers indicating potentially more stable Newton steps. For the cubic spring, the SFN-based Jacobian is compared with the analytical reference and shows almost identical conditioning. 
For the unilateral spring, the effective condition number is considered due to occasional rank deficiencies. The reference nonlinear-force Jacobian vanishes in regions where the contact remains open and changes abruptly as contact becomes active. In contrast, the SFN represents a smooth approximation of the force mapping and therefore generally exhibits small but nonzero local derivatives also outside the sharply defined contact transitions. This discrepancy reflects an approximation of the local force sensitivity rather than an effect of automatic differentiation itself. Numerically, the resulting smoothing can act as a regularization of the Jacobian and may improve its conditioning, although the improved conditioning should not by itself be interpreted as a more accurate Jacobian. In the regions where both Jacobians can be compared, the condition number of the AFT-based Jacobian is several orders of magnitude larger than that of the SFN-based Jacobian. For the Jenkins element, the condition numbers of the AFT- and SFN-based Jacobians follow similar trends along the frequency sweep. However, over parts of the intermediate frequency range, between $\Omega=0.7$ and $1.1$, the SFN-based Jacobian is better conditioned and exhibits lower condition numbers. Outside this range, the differences are smaller, with either formulation occasionally yielding lower condition numbers.

The SFN-based Jacobians also result in generally comparable solver convergence behavior. For the cubic spring, both formulations require the same number of Newton-type iterations at all continuation points. For the Jenkins element, the iteration counts agree at 99.5\% of the points, with the SFN requiring fewer iterations at the remaining 0.5\%. The unilateral spring exhibits larger local differences: the iteration counts are identical at 83.2\% of the continuation points, while the SFN requires fewer iterations at 5.2\%, one additional iteration at 9.8\%, and more than one additional iteration at 1.8\%. Additional SFN iterations occur predominantly in sensitive regions associated with contact transitions and around the resonance peak, where differences in the learned force mapping and its local derivatives have a stronger influence on the nonlinear correction. Conversely, fewer iterations are mainly observed in the closed-contact regime. Although these regions partly coincide with differences in Jacobian conditioning, the condition number alone does not determine the nonlinear convergence rate.

These results show that SFNs can be seamlessly embedded into an existing HBM solver while preserving the relevant convergence and solution behavior for the investigated test cases. Along the regions visited during the frequency continuations, the predicted nonlinear force coefficients and their Jacobians agree closely with the corresponding AFT-based reference. Consequently, the resulting frequency response curves closely reproduce the reference solutions. The Newton iterations remain stable, with convergence behavior comparable to the reference and slight improvements for selected parameter configurations in nonsmooth contact problems.
This favorable convergence behavior is particularly noteworthy because accurate force predictions alone do not guarantee Jacobians that are suitable for Newton-type iterations. Even if the predicted force coefficients remain bounded over the considered input domain, this does not constrain how sensitively the learned mapping responds to small variations of the displacement coefficients. In principle, small input perturbations could therefore lead to disproportionately large changes in the predicted force coefficients or their local derivatives. The fact that such behavior is not observed in the investigated cases indicates that the SFNs capture not only the nonlinear force mapping but also the relevant local derivatives along the considered continuation paths.

\subsection{Applicability under Parameter Variations}
\begin{figure}
    \centering
    \includegraphics[width=0.85\linewidth]{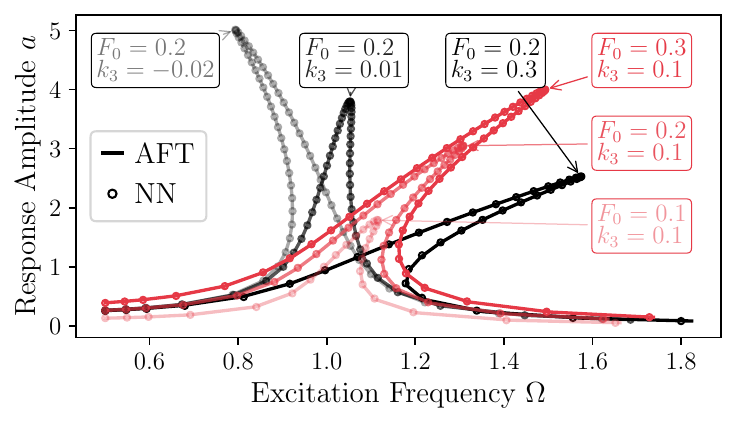}
    \\
    \includegraphics[width=0.88\linewidth]{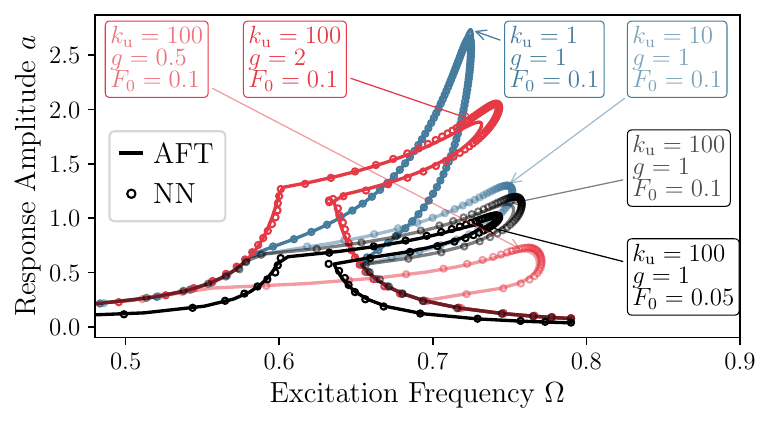}
    \\   
    \includegraphics[width=.9\linewidth]{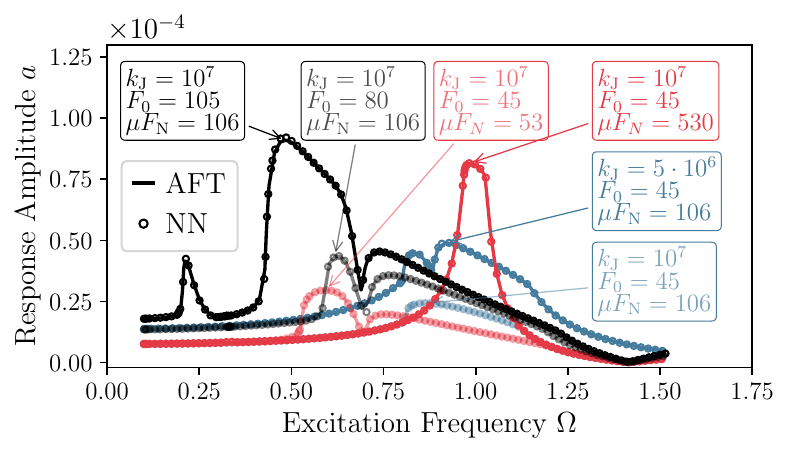}
    \\
    \includegraphics[width=0.88\linewidth]{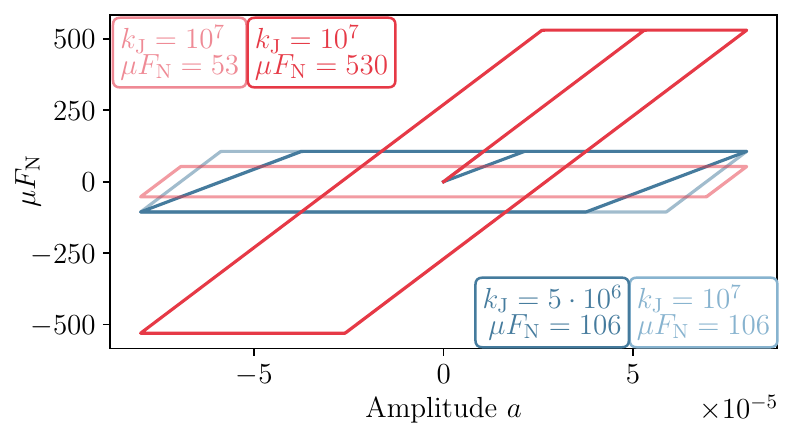} \hspace{0.2em}
    \caption{Frequency response curves for parameter variations of the cubic spring (top), unilateral spring (second), and Jenkins element (third), comparing AFT reference solutions (lines) with SFN predictions (markers). For each nonlinearity, the same trained SFN is used for all parameter sets with independently adaptive continuation. The bottom panel shows the hysteresis loops for the Jenkins-element configurations at a common relative displacement amplitude $a=0.8\times10^{-4}\,\mathrm{m}$, approximately corresponding to the maximum response amplitude of the configuration with the highest friction limit.}
    \label{fig:frc_variations}
\end{figure}

The applicability of the SFNs under variations of the physical parameters of the nonlinear force laws is assessed for parameter configurations exhibiting qualitatively different nonlinear frequency response behavior. In contrast to the preceding fixed-step comparisons, the continuation step size is adapted automatically in this study, allowing the solver to respond freely to local changes in the nonlinear solution behavior. Thereby potential differences in the adaptive step-size selection and nonlinear corrector behavior of the AFT- and SFN-based formulations can be exposed.
The resulting frequency response curves are shown in Figure~\ref{fig:frc_variations}, while detailed continuation statistics, including the number of points on the solution curve $n_{\mathrm{cont}}$, the total and mean number of Newton-type iterations, and the total number of function evaluations, are reported in Appendix~\ref{tab:continuation_statistics}.

Importantly, the parameter variations do not constitute out-of-distribution generalization with respect to the SFN inputs. Through the physics-based preprocessing introduced for each application case, the relevant physical parameters are analytically separated from the mapping learned by the network, while phase normalization removes the redundant dependence on the first-harmonic phase. Consequently, different physical parameter configurations are mapped onto the same parameter-independent SFN formulation, provided that the resulting normalized displacement coefficients remain within the sampled training domain.

The parameter variations are chosen to produce clearly different nonlinear frequency-response behavior and, consequently, different regions of the normalized coefficient space visited by the HBM solver. The purpose of this study is therefore to assess whether the preprocessing and the selected training domains are sufficiently broad to cover these parameter-induced response states. For all considered configurations, the resulting SFN inputs remain within the sampled training ranges.

After training on the domains specified in Appendix~\ref{app:training_data}, the SFN weights are fixed and the same network is used for all parameter configurations shown in Figure~\ref{fig:frc_variations}. Thus, only three SFNs are employed in total: one for the cubic spring, one for the unilateral spring, and one for the Jenkins element.

For the cubic spring, the parameter variations cover both hardening and softening Duffing regimes. Depending on the cubic stiffness and excitation level, the responses range from nearly linear behavior to strongly nonlinear frequency response curves with pronounced resonance bending toward higher frequencies for positive cubic stiffness and toward lower frequencies for negative cubic stiffness. 

For the unilateral spring, the parameter variations cover different contact activation regimes. Varying the excitation level changes the extent to which the gap is exceeded during the oscillation cycle, whereas variations of the gap shift the onset of contact to different response amplitudes. The contact stiffness controls the severity of contact interaction once the unilateral spring is active. As a result, the considered cases range from weakly activated contact responses to strongly nonsmooth frequency response curves with pronounced turning points. 

For the Jenkins element, the parameter variations cover different frictional contact regimes. The excitation amplitude controls the level of activation of the frictional nonlinearity, while the tangential stiffness and friction limit force determine the transition between sticking and sliding. The resulting responses range from nearly sticking behavior with a high effective tangential stiffness to pronounced stick-slip motion with increased hysteretic dissipation. The frequency responses further exhibit a resonant modal interaction, visible as a characteristic double-peak structure whose prominence varies with the considered parameter configuration. The hysteresis loops in the bottom panel of Figure~\ref{fig:frc_variations} additionally illustrate the associated changes in local stick-slip behavior and frictional dissipation.

The resulting frequency response curves demonstrate that, for each application case, the same respective SFN can be reused across all considered physical parameter configurations without retraining. Despite the qualitatively different nonlinear response behavior and the corresponding variation of the coefficient-space regions visited by the HBM solver, the SFN-based solutions closely reproduce the AFT-based reference responses. For the cubic spring, this includes both hardening and softening configurations, enabled by the analytical separation of the cubic stiffness from the parameter-independent mapping learned by the SFN. Likewise, the nondimensional formulations of the unilateral spring and Jenkins element allow variations of their physical force parameters to be represented by the same respective networks. These results therefore demonstrate that the physics-based preprocessing substantially enlarges the range of physical configurations covered by a single trained SFN, provided that the resulting normalized and phase-normalized displacement coefficients remain within the sampled training domain. Parameter variations that drive these coefficients outside this domain constitute genuine extrapolation and are not assumed to be represented reliably without extending the training data.

The adaptive continuation results show that the SFN-based formulation generally preserves the solver behavior across the investigated parameter configurations. The numbers of continuation points, Newton-type iterations, and function evaluations remain comparable to the AFT reference for all three nonlinearities, with only moderate case-dependent deviations. Detailed continuation statistics are provided in Table~\ref{tab:continuation_statistics}.

\subsection{Performance Comparison}\label{sec:performance_comparison}

\begin{figure}
\centering
\includegraphics[width=.98\linewidth]{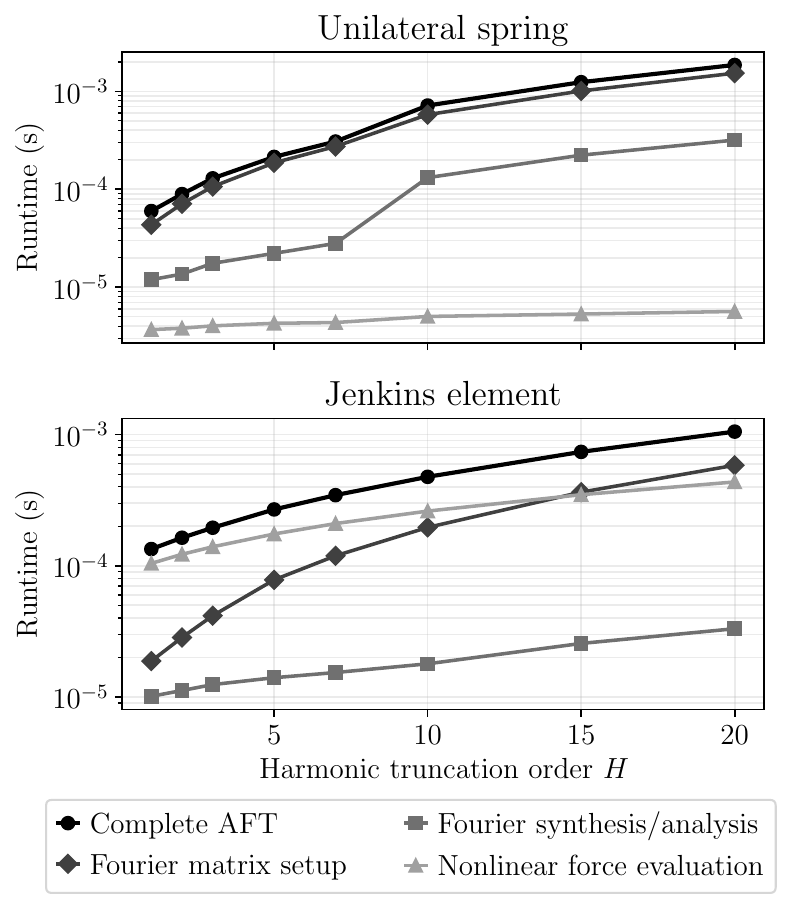}
\caption{Runtime composition of the AFT evaluation as a function of the harmonic truncation order \(H\) for the unilateral spring and Jenkins element.}
\label{fig:aft_cost_composition}
\end{figure}

\begin{figure}
\centering
\includegraphics[width=\linewidth]{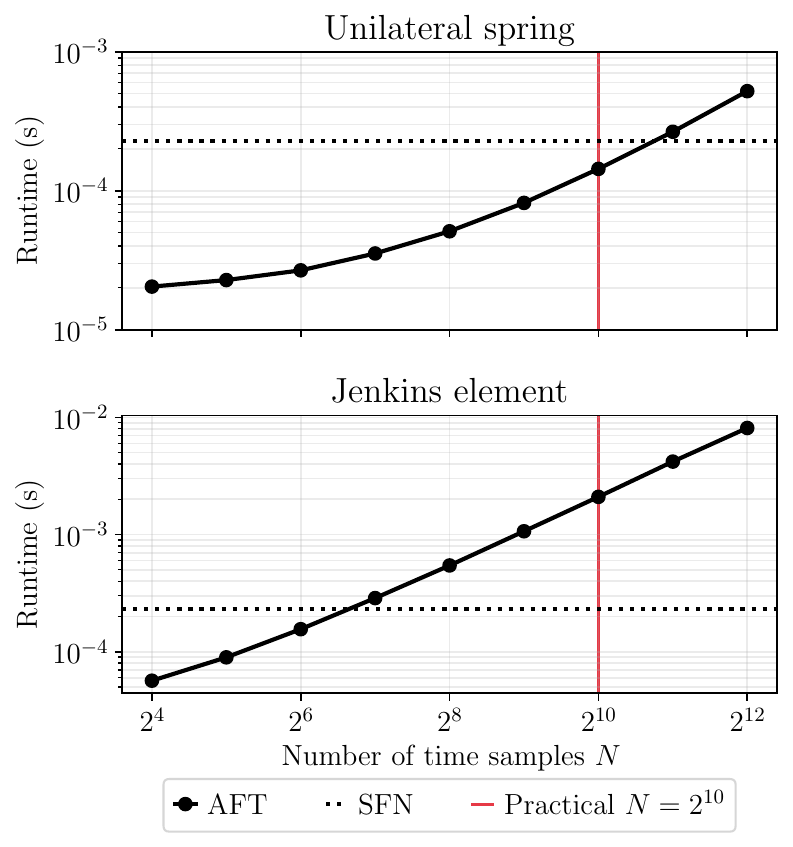}
\caption{Runtime comparison of AFT and SFN evaluations as a function of the number of time samples \(N\) for the unilateral spring and Jenkins element. The practical resolution \(N=2^{10}\) is indicated.}
\label{fig:runtime_resolution}
\end{figure}

To assess the computational performance independently of the surrounding HBM implementation, the Python implementations of the nonlinear-force mappings are benchmarked on a single CPU thread. Python--MATLAB interactions are excluded, and neither Jacobian computation nor the Newton-type solver is considered. The analysis is restricted to the unilateral spring and the Jenkins element, since these represent the nonsmooth and hysteretic cases for which a comparatively high temporal resolution is required. The smooth cubic spring, for which the nonlinear force can be evaluated inexpensively with a low temporal resolution, is therefore not considered in the present performance study. The computations are performed on an x86-64 CPU system with 16 physical cores and $33.8\,\mathrm{GB}$ of memory, while all benchmarked evaluations are restricted to a single CPU thread.

First, the practical composition of the AFT runtime is investigated under variation of the harmonic truncation order $H$. For each $H$, the number of time samples $N$ is selected according to the empirical resolution rules $N_{\mathrm{emp}}(H)=\mathrm{min}(500+25\cdot H, 2000)$ for the unilateral spring and $N_{\mathrm{emp}}(H)=\mathrm{min}(50+10\cdot H,500)$ for the Jenkins element given in~\cite{woiwode_2020}. This accounts for the practical coupling between harmonic truncation and temporal resolution. Figure~\ref{fig:aft_cost_composition} compares the runtime of the complete AFT evaluation with the contributions of the Fourier-matrix setup, Fourier synthesis and analysis, and the nonlinear force evaluation. The results illustrate that the asymptotic scaling of the individual operations alone does not determine their practical runtime contribution. For the unilateral spring, the construction of the Fourier transformation matrix constitutes the dominant part of the complete AFT evaluation. The actual Fourier synthesis and analysis operations as well as the pointwise nonlinear force evaluation contribute only a comparatively small fraction. For the Jenkins element, the Fourier-matrix setup remains relevant, but the sequential hysteretic force evaluation becomes the dominant contribution. In contrast, the actual Fourier synthesis and analysis operations remain subordinate over the investigated range. These observations confirm that practical bottlenecks cannot generally be inferred from asymptotic complexity estimates alone and depend strongly on the nonlinear force law and its implementation.

Second, the AFT and SFN runtimes are compared directly as a function of the temporal resolution $N$, while the harmonic truncation order is fixed to the value for which the respective SFN was trained, namely $H=2$ for the unilateral spring and $H=3$ for the Jenkins element. Since the SFN operates directly on the retained Fourier coefficients, its inference cost is independent of $N$, whereas the AFT runtime increases with temporal resolution. Figure~\ref{fig:runtime_resolution} therefore also illustrates the resolution at which SFN inference becomes computationally advantageous.

For the unilateral spring, the AFT remains faster than the SFN at the practically employed resolution of $N=2^{10}$. Only at higher temporal resolutions does the increasing AFT cost lead to a crossover in favor of the SFN. This is consistent with the inexpensive, pointwise evaluation of the unilateral force law. For the Jenkins element, in contrast, the SFN becomes advantageous already at comparatively low temporal resolutions. At the practically employed resolution of $N=2^{10}$, its inference time is approximately one order of magnitude smaller than that of the AFT. This advantage originates from avoiding the sequential time marching required to establish and evaluate the hysteretic force history over multiple periods. The comparison therefore indicates that the computational benefit of the SFN is particularly pronounced for history-dependent nonlinearities, while for inexpensive memoryless force laws it depends on the temporal resolution required by the AFT.

\section{Conclusion}\label{sec:discussion}
The AFT scheme provides a general means of evaluating nonlinear forces within the HBM, but its computational cost increases with the temporal resolution required for the nonlinear force evaluation. This becomes particularly relevant for hysteretic force laws requiring sequential time-domain evaluation and suitable initialization of internal states. In this work, these costs are addressed by replacing the AFT evaluation of individual nonlinear elements with differentiable SFNs operating directly on their Fourier coefficients.

The results for smooth, nonsmooth, and hysteretic nonlinearities demonstrate that the SFNs can be integrated into the existing solver and continuation framework while replacing only the AFT evaluation. Despite the larger approximation errors for the nonsmooth unilateral spring, the resulting frequency response curves reproduce the overall AFT-based solution branches and resonance characteristics, with local deviations occurring primarily in sensitive regions. The numbers of Newton-type iterations required by the SFNs are identical or lower for 100\%, 88.4\%, and 100\% of the continuation points for the three application cases, respectively, showing that the learned force mappings and their AD-based Jacobians generally preserve the numerical behavior of the original formulation.

Physics-based parameter separation, nondimensionalization, and phase normalization remove explicit parameter dependencies and redundant phase information from the learned mappings. Consequently, a single trained SFN can be reused across a wide range of physical parameter configurations as long as the resulting normalized coefficient states are covered by the training domain. The present results further show that this domain can either be sampled broadly to promote general applicability or restricted using application-specific physical knowledge to improve accuracy and reduce the required training data. While an individual SFN remains specific to a nonlinear force law and harmonic truncation order, this formulation provides a basis for a reusable library of spectral nonlinear-element models that can be assembled modularly within larger mechanical systems.

The main computational potential arises for problems requiring high temporal resolution or containing many nonlinear elements. For a fixed SFN, inference and AD-based differentiation are independent of the number of AFT time samples and support vectorized batch evaluation. Consequently, substantial acceleration is not necessarily expected for inexpensive nonlinearities at low temporal resolution, but becomes increasingly relevant for history-dependent force evaluations and large numbers of nonlinear interfaces. A modular library of reusable SFNs therefore provides a path toward scalable high-fidelity nonlinear response analyses, parameter studies, uncertainty quantification, and design optimization while retaining close agreement with established AFT-based HBM solutions.

\printcredits

\section*{Funding}
This work was funded by the German Federal Ministry for Economic Affairs and Energy (BMWE), Grant No. 03EE5189B, and co-funded by FVV e.V., Grant No. 601550.

\section*{Declaration of generative AI and AI-assisted technologies}
During the preparation of this work, the authors used OpenAI ChatGPT (GPT-5) to improve the wording and clarity of selected passages and to assist with the presentation and consistency checking of mathematical expressions. GitHub Copilot was used to a limited extent for routine coding tasks, including documentation, refactoring, assistance with plotting code, and the correction of minor coding errors. All AI-assisted outputs were critically reviewed and edited and, where applicable, tested and verified by the authors. The authors take full responsibility for the content of the publication.

\section*{Data Availability Statement}
The software and data supporting the findings of this study will be made publicly available at \newline \href{https://github.com/MiriamAlina/spectral-force-network/}{https://github.com/MiriamAlina/spectral-force-network/}.

\bibliographystyle{\bstsource}

\bibliography{\bibsource}

\appendix

\section{Problem-Specific Scaling and Parameter Separation}
\subsection{Cubic Spring}\label{app:duffing_parameter_separation}
For the cubic spring with force law $f_{\mathrm{nl}}(t)=k_3 q(t)^3$, the stiffness coefficient $k_3$ enters the nonlinear force only as a scalar factor. Since the Fourier operator is linear, this factor can be separated from the AFT scheme as
\begin{equation}
\begin{split}
    \hat{\mathbf f}_{\mathrm{nl},e,H}^{\mathrm{AFT}}
    &= \mathcal F \left[k_3 \left( \mathcal F^{-1} \left(\hat{\mathbf q}_{e,H}\right) \right)^3 \right] \\
    &= k_3\, \underbrace{\mathcal F \left[ \left( \mathcal F^{-1} \left(\hat{\mathbf q}_{e,H}\right) \right)^3 \right] }_{\overset{!}{\approx} f_{\theta}(\hat{\mathbf{q}}_{e,H})} \quad .
\end{split}
\end{equation}
Here, $\mathcal{F}^{-1}$ and $\mathcal{F}$ denote the discrete Fourier synthesis and analysis operators, respectively. The notation is independent of their numerical implementation, which can use either the explicit transformation matrices introduced above or FFT-based algorithms.

Thus, the SFN maps the Fourier coefficients of the displacement to the retained Fourier coefficients of $q^3$, independently of the value and sign of $k_3$.

The nonlinear force coefficients are then recovered according to
\begin{equation}
    \hat{\mathbf f}_{\mathrm{nl},e,H}^{\mathrm{SFN}} =  k_3 f_{\theta}\left(\hat{\mathbf q}_{e,H}\right) \quad .
\end{equation}
Since $k_3$ is independent of the displacement coefficients, the corresponding Jacobian follows directly as
\begin{equation}
    \frac{\partial \hat{\mathbf f}_{\mathrm{nl},e,H}^{\mathrm{SFN}}} {\partial\hat{\mathbf q}_{e,H}} = k_3 \frac{\partial f_{\theta}\left(\hat{\mathbf q}_{e,H}\right)} {\partial\hat{\mathbf q}_{e,H}} \quad .
\end{equation}

The same trained network can therefore be used for arbitrary positive and negative values of $k_3$, representing hardening and softening behavior, respectively. This generalization remains valid as long as the resulting displacement coefficients lie within the input domain represented in the training data.

\subsection{Unilateral Spring}\label{app:unilateral_scaling}
Introducing the dimensionless displacement and force
\begin{equation}
    q^*(t)=\frac{q(t)}{g}, \qquad
    f_{\mathrm{nl}}^*(t) = \frac{f_{\mathrm{nl}}(t)}{k_{\mathrm u}g} \quad ,
\end{equation}
yields the parameter-independent force law
\begin{equation}
    f_{\mathrm{nl}}^*(t) = \bigl(q^*(t)-1\bigr) \mathbf{1}_{q^*(t)\geq1} \quad .
\end{equation}

The Fourier coefficient vectors are scaled accordingly, such that the SFN approximates the dimensionless mapping
\begin{equation}
    \hat{\mathbf q}_{e,H}^* = \frac{\hat{\mathbf q}_{e,H}}{g} \quad , \qquad
    f_{\theta}\left(\hat{\mathbf q}_{e,H}^*\right) \approx\hat{\mathbf f}_{\mathrm{nl},e,H}^* = \frac{\hat{\mathbf f}_{\mathrm{nl},e,H}} {k_{\mathrm u}g} \quad .
\end{equation}
The nonlinear force coefficients in physical units are then recovered as
\begin{equation}
    \hat{\mathbf f}_{\mathrm{nl},eH}^{\mathrm{SFN}} = k_{\mathrm u}g\, f_{\theta}\left(\hat{\mathbf q}_{e,H}^*\right) \quad .
\end{equation}

Since $\frac{\partial \hat{\mathbf{q}}_{e,H}^*}{\partial \hat{\mathbf{q}}_{e,H}} = \frac{1}{g} \mathbf{I}$, the corresponding Jacobian in physical units follows from the chain rule as
\begin{equation}
    \frac{\partial \hat{\mathbf f}_{\mathrm{nl},e,H}^{\mathrm{SFN}}} {\partial\hat{\mathbf q}_{e,H}} = k_{\mathrm u}
    \frac{\partial f_{\theta}\left(\hat{\mathbf q}_{e,H}^*\right)} {\partial\hat{\mathbf q}_{e,H}^*} \quad .
\end{equation}

\subsection{Jenkins Element}\label{app:jenkins_scaling}

The transition between sticking and sliding is governed by the ratio of elastic spring force $k_{\mathrm{J}} q$ to friction limit force $\mu F_N$. The dimensionless displacement and force are defined as 
\begin{equation}
    q(t)^* = \frac{k_{\mathrm{J}} q(t)}{\mu F_\mathrm{N}} \quad , \quad f_{\mathrm{nl}}(t)^* = \frac{f_\mathrm{nl}(t)}{\mu F_{\mathrm{N}}} \quad .
\end{equation}

The time-discrete Jenkins update then becomes
\begin{equation}
    f_{\mathrm{nl},i}^* = \operatorname{clip} \left( f_{\mathrm{nl},i-1}^* + q_i^*-q_{i-1}^*, -1,1 \right) \quad .
\end{equation}
This nondimensional representation removes the explicit dependence on $k_{\mathrm{J}}$ and $\mu F_{\mathrm{N}}$ from the core hysteretic mapping and the dimensionless nonlinear force response follows a universal curve. This normalization is motivated by the scale invariance of piecewise linear contact constraints discussed in~\cite{krack_reliability_2014, Krack2024} and allows responses with different contact stiffness and friction limit to be represented on a common scale. 

The Fourier coefficient vectors are scaled accordingly, such that the SFN approximates only the core mapping
\begin{equation}
    \hat{\mathbf q}_{e,H}^* = \frac{k_{\mathrm J}}{\mu F_{\mathrm{N}}}\hat{\mathbf q}_{e,H} \quad , \qquad
    f_{\theta}\left(\hat{\mathbf q}_{e,H}^*\right) \approx \hat{\mathbf f}_{\mathrm{nl},e,H}^* = \frac{\hat{\mathbf f}_{\mathrm{nl},e,H}}{\mu F_{\mathrm{N}}} \quad .
\end{equation}
The nonlinear force coefficients in physical units are recovered as
\begin{equation}
    \hat{\mathbf f}_{\mathrm{nl},e,H}^{\mathrm{SFN}} = \mu F_{\mathrm{N}} f_{\theta}\left(\hat{\mathbf q}_{e,H}^*\right).
\end{equation}
Since $\frac{\partial\hat{\mathbf q}_{e,H}^*} {\partial\hat{\mathbf q}_{e,H}} = \frac{k_{\mathrm J}}{\mu F_{\mathrm{N}}}\mathbf I$, the corresponding Jacobian follows from the chain rule as
\begin{equation}
    \frac{\partial \hat{\mathbf f}_{\mathrm{nl},e,H}^{\mathrm{SFN}}} {\partial\hat{\mathbf q}_{e,H}} = k_{\mathrm J} \frac{\partial f_{\theta}(\hat{\mathbf q}_{e,H}^*)} {\partial\hat{\mathbf q}_{e,H}^*}.
\end{equation}

\section{Phase Normalization}
\label{app:phase_normalization}

When the nonlinear force law has no explicit time dependence, its mapping is equivariant with respect to time shifts: a phase shift of the displacement signal produces the corresponding harmonic-wise phase shift of the nonlinear force. For history-dependent force laws, phase equivariance applies to the established periodic state: the input history, internal state, and resulting force history are shifted together.

Therefore, the phase
\begin{equation}
    \phi(\hat{\mathbf q}_{e,H})
    =
    \operatorname{atan2}(-b_1,a_1)
\end{equation}
can be removed before evaluating the SFN. It is chosen such that the fundamental harmonic becomes a pure cosine.

The phase-normalized displacement coefficients are obtained as
\begin{equation}
    \hat{\mathbf q}'_{e,H}
    =
    \mathbf R_{e,H}\!\left(\phi(\hat{\mathbf q}_{e,H})\right)
    \hat{\mathbf q}_{e,H} \quad ,
\end{equation}
where
\begin{equation}
    \mathbf R_{e,H}(\phi)
    =
    \operatorname{blkdiag}
    \left(
        1,\mathbf R_{e,1}(\phi),\ldots,\mathbf R_{e,H}(\phi)
    \right)
\end{equation}
with
\begin{equation}
    \mathbf R_{e,h}(\phi)
    =
    \begin{bmatrix}
        \cos(h\phi) & -\sin(h\phi)\\
        \sin(h\phi) &  \cos(h\phi)
    \end{bmatrix} \quad .
\end{equation}
In particular,
\begin{equation}
    a'_1=\sqrt{a_1^2+b_1^2} \quad ,
    \qquad
    b'_1=0 \quad .
\end{equation}
The normalization removes redundant phase information and prevents the SFN from having to learn the corresponding rotational equivariance from the training data.

The predicted nonlinear force coefficients are transformed back to the original phase according to
\begin{equation}
    \hat{\mathbf f}_{\mathrm{nl},e,H}^{\mathrm{SFN}}
    =
    \mathbf R_{e,H}\!\left(-\phi(\hat{\mathbf q}_{e,H})\right)
    f_{\theta}\!\left(\hat{\mathbf q}'_{e,H}(\hat{\mathbf q}_{e,H})\right) \quad .
\end{equation}
Applying the chain rule gives the corresponding Jacobian,
\begin{equation}
\begin{split}
    \frac{\partial
    \hat{\mathbf f}_{\mathrm{nl},e,H}^{\mathrm{SFN}}}
    {\partial\hat{\mathbf q}_{e,H}}
    ={}&
    \mathbf R_{e,H}(-\phi)
    \frac{\partial f_{\theta}(\hat{\mathbf q}'_{e,H})}
    {\partial\hat{\mathbf q}'_{e,H}}
    \frac{\partial\hat{\mathbf q}'_{e,H}}
    {\partial\hat{\mathbf q}_{e,H}}
    \\[0.3em]
    &+
    \left[
        \frac{\partial\mathbf R_{e,H}(-\phi)}
        {\partial\phi}
        f_{\theta}(\hat{\mathbf q}'_{e,H})
    \right]
    \frac{\partial\phi}
    {\partial\hat{\mathbf q}_{e,H}} \quad .
\end{split}
\end{equation}
The derivative of the normalized input is
\begin{equation}
    \frac{\partial\hat{\mathbf q}'_{e,H}}
    {\partial\hat{\mathbf q}_{e,H}}
    =
    \mathbf R_{e,H}(\phi)
    +
    \left[
        \frac{\partial\mathbf R_{e,H}(\phi)}
        {\partial\phi}
        \hat{\mathbf q}_{e,H}
    \right]
    \frac{\partial\phi}
    {\partial\hat{\mathbf q}_{e,H}} \quad .
\end{equation}
For
$\hat{\mathbf q}_{e,H}=[a_0,a_1,b_1,\ldots,a_H,b_H]^\top$,
the phase derivative is
\begin{equation}
    \frac{\partial\phi}
    {\partial\hat{\mathbf q}_{e,H}}
    =
    \begin{bmatrix}
        0 &
        \dfrac{b_1}{a_1^2+b_1^2} &
        -\dfrac{a_1}{a_1^2+b_1^2} &
        0 & \cdots & 0
    \end{bmatrix} \quad .
\end{equation}
The normalization is well defined for a nonzero fundamental-harmonic amplitude. If \(a_1^2+b_1^2\) approaches zero, the phase becomes undefined and must be treated separately.

\section{Training Data}\label{app:training_data}
The training data are generated directly in the phase-normalized coefficient space introduced for the respective application cases. Coefficients marked by $(\cdot)^*$ are additionally nondimensionalized according to the corresponding physics-based scaling. All input coefficients are sampled independently from uniform distributions, and the corresponding nonlinear force coefficients are obtained by AFT and used as training targets. The sampling domains and dataset sizes are summarized in Table~\ref{tab:training_domains}, where $\mathcal U[l,u]$ denotes a uniform distribution over $[l,u]$.

\begin{table}[]
    \centering
    \caption{Sampling domains and dataset sizes used for the generation of the SFN training data.}
    \label{tab:training_domains}
    \begin{tabular}{llll}
        \toprule
        Test case & Samples & Coefficient & Sampling rule \\
        \midrule
        \multirow{3}{*}{Cubic spring} & \multirow{3}{*}{$10^5$} & $a_1'$ & $\mathcal{U} [0,\,5]$ \\
                        &         & $a_3'$ & $\mathcal{U} [-5,\,5]$ \\
                        &         & $b_3'$ & $\mathcal{U} [-5,\,5]$ \\
        \midrule
        \multirow{4}{*}{Unilateral spring} & \multirow{4}{*}{$10^6$} & $a_0^{*,\prime}$ & $\mathcal{U} [-2,\,2]$ \\
                        &         & $a_1^{*,\prime}$ & $\mathcal{U} [0,\,2]$ \\
                        &         & $a_2^{*,\prime}$ & $\mathcal{U} [-2,\,2]$ \\
                        &         & $b_2^{*,\prime}$ & $\mathcal{U} [-2,\,2]$ \\
        \midrule
        \multirow{3}{*}{Jenkins element} & \multirow{3}{*}{$10^6$} & $a_1^{*,\prime}$ & $\mathcal{U}[0,\,5]$ \\
        & & $a_3^{*,\prime}$ & $\mathcal{U}[-5,\,5]$ \\
        & & $b_3^{*,\prime}$ & $\mathcal{U}[-5,\,5]$ \\
        \bottomrule
    \end{tabular}
\end{table}

\section{Physics-Informed Training Data Sampling for the Unilateral Spring}\label{app:unilateral_physics_sampling}
While the main training data are sampled broadly without application-specific assumptions, additional physical information can be used to restrict the relevant coefficient space. For the considered SDOF system with the unilateral spring as the only nonlinear element, open contact implies a linear response under zero-mean first-harmonic excitation, such that \(a_0^{*,\prime}=a_2^{*,\prime}=b_2^{*,\prime}=0\). Moreover, the higher-harmonic coefficients remain small after contact activation, allowing the sampling to be concentrated accordingly. The resulting physics-informed sampling strategy is summarized in Table~\ref{tab:unilateral_sampling}.

\begin{table}[]
    \centering
    \caption{Application-specific sampling strategy for unilateral spring training data.}
    \label{tab:unilateral_sampling}
    \begin{tabular}{llll}
        \toprule
        Regime & Samples & Coeff. & Sampling rule \\
        \midrule
        \multirow{4}{*}{\makecell{Open/\\onset}} & \multirow{4}{*}{2500} & $a_0^{*,\prime}$ & 0 \\
        & & $a_1^{*,\prime}$ & $\mathcal{N}_{[0.01,\,1.1]} (1.0,\,0.3)$ \\
        & & $a_2^{*,\prime}$ & 0 \\
        & & $b_2^{*,\prime}$ & 0 \\
        \midrule
        \multirow{7}{*}{\makecell{Contact-\\active}} & \multirow{7}{*}{7500} & $a_0^{*,\prime}$ & 
        $
        \begin{aligned}
            \mathcal{N}_{[-0.2,\,0.2]} (-0.05,\,0.04) \\ - 0.6(a_1^{*,\prime} -1)
        \end{aligned}
        $
        \\
        & & $a_1^{*,\prime}$ & $\mathcal{N}_{[1.0,\,2.0]} (1.5,\,0.3)$ \\
        & & $a_2^{*,\prime}$ 
        & 
        $
        \begin{aligned}
            \mathcal{N}_{[-0.2,\,0.2]} (0.15,\,0.04) \\
            - 0.5(a_1^{*,\prime} -1)
        \end{aligned}
        $
         \\
        & & $b_2^{*,\prime}$ & 
        $
        \begin{aligned}
        \mathcal{N}_{[-0.05,\,0.05]} (0,\,0.01) \\
        - 0.09(a_1^{*,\prime} -1)
        \end{aligned}
        $ \\
        \bottomrule
    \end{tabular}
\end{table}

\begin{figure}
    \centering
    \includegraphics[width=\linewidth]{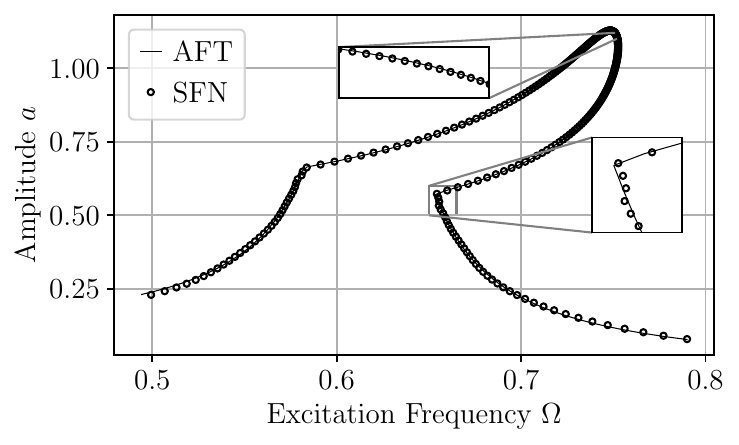}
    \caption{Frequency-response curve obtained with an SFN for the unilateral spring that was trained on physics-informed training data.}
    \label{fig:unilateral_frc_pi_sampling}
\end{figure}

Incorporating these application-specific constraints substantially increases the density of training samples in the region of coefficient space actually visited by the frequency-response solution. As shown in Fig.~\ref{fig:unilateral_frc_pi_sampling}, this leads to a markedly improved agreement around the resonance peak and allows comparable or improved SFN accuracy to be obtained with a substantially smaller number of training samples of $10^4$ than for the more general sampling strategy with $10^6$ training samples.

This illustrates the trade-off between generality and data efficiency. Application-specific physical knowledge can improve accuracy and reduce training effort at the expense of transferability to more general system configurations.

\section{Neural Network Specifications}\label{app:NN_specifications}
A separate SFN is trained for each nonlinear force law and for a fixed harmonic truncation order per model. All SFNs are fully connected feed-forward neural networks with GELU activations in the hidden layers and a linear output layer. Prior to training, the physically preprocessed input and output coefficients are standardized using the mean and standard deviation of the respective training data. This statistical standardization is applied in addition to phase normalization and, where applicable, case-specific nondimensionalization.

The network architectures are summarized in Table~\ref{tab:NN_specifications}. The cubic-spring SFN uses three phase-normalized input coefficients, whereas the unilateral-spring and Jenkins-element SFNs operate on four and three nondimensionalized and phase-normalized inputs, respectively. The corresponding output dimensions follow from the retained nonlinear force coefficients. The network size is increased for the nonsmooth and hysteretic mappings compared with the smooth cubic-spring case.

All networks are trained by minimizing the mean squared error between the SFN predictions and the corresponding AFT-generated target coefficients using the Adam optimizer.

\begin{table*}
    \centering
    \caption{Architecture and training strategy of the SFNs used for the three application cases.}
    \label{tab:NN_specifications}
    \begin{tabular}{lccc}
        \toprule
        & Cubic spring & Unilateral spring & Jenkins element \\
        \midrule
        Network type                       & Feedforward neural network & Feedforward neural network & Feedforward neural network \\
        Input dimension                    & 3 & 4 & 3 \\
        Output dimension                   & 4 & 5 & 4 \\
        Hidden layers                      & 3 & 5 & 5 \\
        Neurons per hidden layer           & 128 & 128 & 128 \\
        Activation                         & GELU & GELU & GELU \\
        Trainable parameters               & 34\,052 & 67\,333 & 67\,076 \\
        \midrule
        Train/validation/test split        & 60/20/20 & 60/20/20 & 60/20/20 \\
        Batch size                         & 128 & 128 & 128 \\
        Learning rate                      & 0.002 & 0.0005 & 0.002 \\
        Loss function
        &
        \begin{tabular}[t]{@{}c@{}}
            MSE \\
            $\displaystyle
                \mathcal{L}_\text{MSE}
                =
                \frac{1}{N}
                \sum_{i=1}^{N}
                (\hat{y}_i-y_i)^2$
        \end{tabular}
        &
        \begin{tabular}[t]{@{}c@{}}
            Zero-weighted MSE \\[0.1em]
            $
            \begin{aligned}[t]
                \mathcal{L}_{\mathrm{zwMSE}}
                &=
                \frac{
                \sum_{i=1}^{N} w_i
                (\hat{y}_i-y_i)^2
                }{
                \sum_{i=1}^{N} w_i
                },\\
                w_i
                &=
                \begin{cases}
                    w_0, & |y_i|\leq\varepsilon,\\
                    1, & \text{otherwise},
                \end{cases} \\
                w_0 &= 200, \\
                \varepsilon &= 10^{-12}
            \end{aligned}
            $
        \end{tabular}
        &
        \begin{tabular}[t]{@{}c@{}}
            MSE \\
            $\displaystyle
                \mathcal{L}_\text{MSE}
                =
                \frac{1}{N}
                \sum_{i=1}^{N}
                (\hat{y}_i-y_i)^2$
        \end{tabular}
        \\
        \bottomrule
    \end{tabular}
\end{table*}

\section{Error Metrics}\label{app:error_metrics}

\begin{figure}
    \centering
    \includegraphics[width=\linewidth]{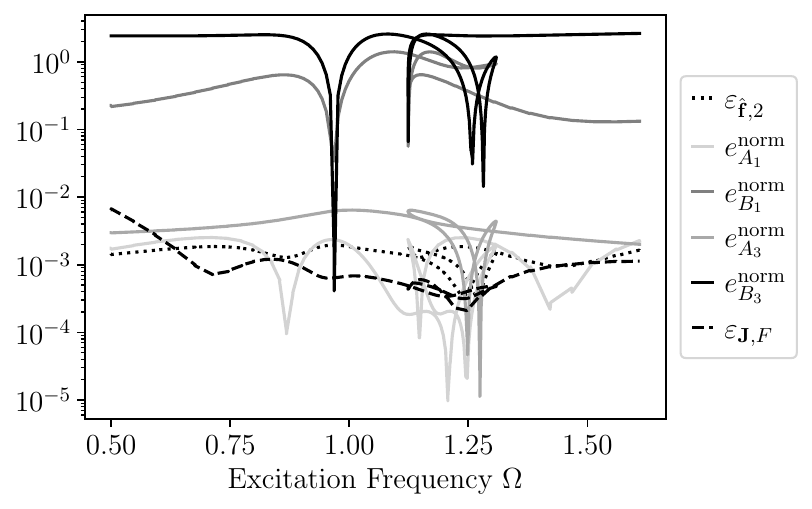}
    \\
    \includegraphics[width=\linewidth]{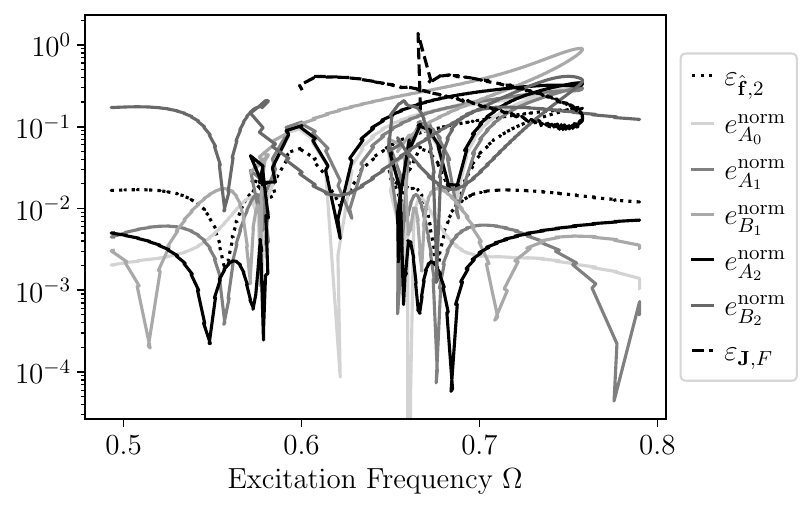}
    \\
    \includegraphics[width=\linewidth]{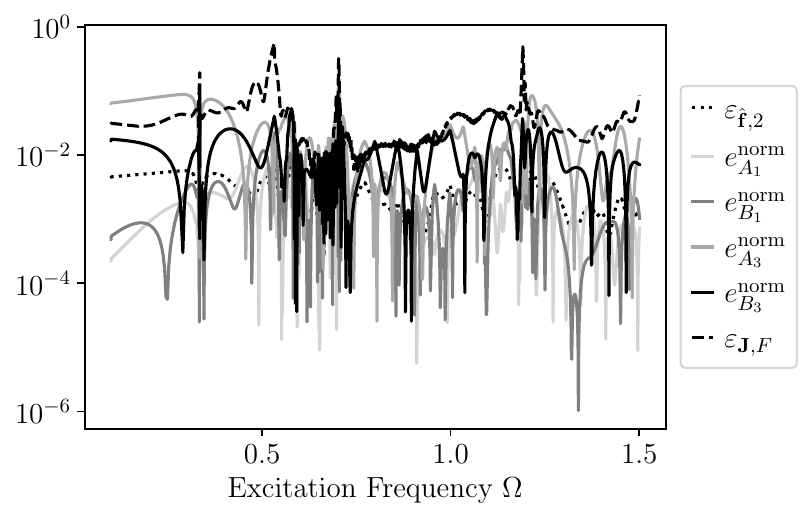}
    \caption{Frequency-resolved force-coefficient and Jacobian errors along the considered frequency-response continuations. Shown are the pointwise normalized \(L^2\)-error of the complete force-coefficient vector \(\varepsilon_{\hat{\mathbf f},2,i}\), the component-wise normalized absolute errors \(e_{l,i}^{\mathrm{norm}}\), and the pointwise relative Frobenius norm error of the Jacobian \(\varepsilon_{\mathbf J,F,i}\).}
    \label{fig:error_metrics_vs_omega}
\end{figure}

All error metrics are evaluated at the same \(n_\Omega\) points on the solution curve using identical displacement-coefficient inputs for the AFT and SFN evaluations. The aggregate metrics reported in Table~\ref{tab:results} quantify the approximation accuracy over the complete continuation path. In addition, frequency-resolved metrics are considered to identify local variations of the approximation error along the frequency response.

\paragraph{Global relative \(L^2\)-error of the force coefficients.}
The overall error of the nonlinear force-coefficient vectors is quantified by
\begin{equation}
    \varepsilon_{\hat{\mathbf f},2}^{\mathrm{glob}} = \sqrt{ \frac{ \displaystyle \sum_{i=1}^{n_\Omega} \left\| \hat{\mathbf f}_{\mathrm{nl},e,H,i}^{\mathrm{AFT}} - \hat{\mathbf f}_{\mathrm{nl},e,H,i}^{\mathrm{SFN}} \right\|_2^2 }{ \displaystyle \sum_{i=1}^{n_\Omega} \left\| \hat{\mathbf f}_{\mathrm{nl},e,H,i}^{\mathrm{AFT}} \right\|_2^2 } } \quad .
    \label{eq:rel_l2}
\end{equation}
This corresponds to the relative \(L^2\)-norm of the force-coefficient error accumulated over the complete continuation path.

To resolve the force-coefficient error locally along the frequency response, a constant reference scale is defined as
\begin{equation}
    f_{\mathrm{ref,rms}} = \sqrt{ \frac{1}{n_\Omega} \sum_{i=1}^{n_\Omega} \left\| \hat{\mathbf f}_{\mathrm{nl},e,H,i}^{\mathrm{AFT}} \right\|_2^2 } \quad .
    \label{eq:force_ref_rms}
\end{equation}
The pointwise normalized \(L^2\)-error is then given by
\begin{equation}
    \varepsilon_{\hat{\mathbf f},2,i} = \frac{ \left\| \hat{\mathbf f}_{\mathrm{nl},e,H,i}^{\mathrm{AFT}} - \hat{\mathbf f}_{\mathrm{nl},e,H,i}^{\mathrm{SFN}} \right\|_2 }{ f_{\mathrm{ref,rms}} } \quad .
    \label{eq:pointwise_l2}
\end{equation}
Using a constant normalization avoids artificially increasing the relative error in regions where the magnitude of the reference force coefficients approaches zero. The global and pointwise measures are related by
\begin{equation}
    \varepsilon_{\hat{\mathbf f},2}^{\mathrm{glob}} = \sqrt{ \frac{1}{n_\Omega} \sum_{i=1}^{n_\Omega} \varepsilon_{\hat{\mathbf f},2,i}^2 } \quad .
\end{equation}

\paragraph{Component-wise normalized errors.}
To compare the approximation accuracy of individual force coefficients with different magnitudes, each coefficient is normalized by its standard deviation along the AFT reference continuation. For coefficient component \(l\),
\begin{equation}
    \bar f_l^{\mathrm{AFT}} = \frac{1}{n_\Omega} \sum_{i=1}^{n_\Omega} \left[ \hat{\mathbf f}_{\mathrm{nl},e,H,i}^{\mathrm{AFT}} \right]_l
\end{equation}
and
\begin{equation}
    \sigma_l^{\mathrm{AFT}} = \sqrt{ \frac{1}{n_\Omega} \sum_{i=1}^{n_\Omega} \left( \left[ \hat{\mathbf f}_{\mathrm{nl},e,H,i}^{\mathrm{AFT}} \right]_l - \bar f_l^{\mathrm{AFT}} \right)^2 } \quad .
\end{equation}
The component-wise normalized RMSE reported in Table~\ref{tab:results} is
\begin{equation}
\begin{split}
    \mathrm{NRMSE}_l = \sqrt{ \frac{1}{n_\Omega} \sum_{i=1}^{n_\Omega} \left( \frac{ \left[ \hat{\mathbf f}_{\mathrm{nl},e,H,i}^{\mathrm{AFT}} \right]_l - \left[ \hat{\mathbf f}_{\mathrm{nl},e,H,i}^{\mathrm{SFN}} \right]_l }{\sigma_l^{\mathrm{AFT}} } \right)^2 } \quad , \\ 
    l=0,\ldots,2H \quad .
\end{split}\label{eq:nrmse}
\end{equation}
Correspondingly, the frequency-resolved component-wise normalized absolute error is
\begin{equation}
    e_{l,i}^{\mathrm{norm}} = \frac{ \left| \left[ \hat{\mathbf f}_{\mathrm{nl},e,H,i}^{\mathrm{AFT}} \right]_l - \left[ \hat{\mathbf f}_{\mathrm{nl},e,H,i}^{\mathrm{SFN}} \right]_l \right| }{ \sigma_l^{\mathrm{AFT}} } \quad .
    \label{eq:pointwise_component_error}
\end{equation}
Thus,
\begin{equation}
    \mathrm{NRMSE}_l = \sqrt{ \frac{1}{n_\Omega} \sum_{i=1}^{n_\Omega} \left(e_{l,i}^{\mathrm{norm}}\right)^2 } \quad .
\end{equation}

\paragraph{Relative Frobenius norm error of the Jacobian.}
The pointwise relative Jacobian error is quantified by
\begin{equation}
    \varepsilon_{\mathbf J,F,i} = \frac{ \left\| \mathbf J_{\mathrm{nl},e,H,i}^{\mathrm{ref}} - \mathbf J_{\mathrm{nl},e,H,i}^{\mathrm{SFN}} \right\|_F }{ \max\left( \left\| \mathbf J_{\mathrm{nl},e,H,i}^{\mathrm{ref}} \right\|_F, \epsilon_J \right) } \quad ,
    \label{eq:pointwise_frob_norm}
\end{equation}
where \(\epsilon_J\) is a small numerical threshold preventing division by zero. The mean pointwise relative Frobenius norm error reported in Table~\ref{tab:results} is
\begin{equation}
    \bar{\varepsilon}_{\mathbf J,F} = \frac{1}{n_\Omega} \sum_{i=1}^{n_\Omega} \varepsilon_{\mathbf J,F,i} \quad .
    \label{eq:frob_norm}
\end{equation}
Here,
\begin{equation}
    \mathbf J_{\mathrm{nl},e,H} = \frac{\partial \hat{\mathbf f}_{\mathrm{nl},e,H}} {\partial \hat{\mathbf q}_{e,H}} \quad .
\end{equation}
Depending on the considered test case, \(\mathbf J_{\mathrm{nl},e,H}^{\mathrm{ref}}\) denotes either the closed-form analytical Jacobian in the frequency domain for the cubic spring or a semi-analytical AFT-based Jacobian obtained by transforming piecewise analytical time-domain derivatives into the frequency domain for the Jenkins element and the unilateral spring.
\\

To assess how the local approximation errors evolve along the frequency-response branches, the pointwise force- and Jacobian-error metrics are plotted against the excitation frequency in Fig.~\ref{fig:error_metrics_vs_omega}.

\section{Adaptive Continuation Solver Statistics}
\label{app:continuation_statistics}

Table~\ref{tab:continuation_statistics} summarizes the solver statistics obtained with adaptive continuation for the AFT- and SFN-based formulations.
Here, $n_{\mathrm{cont}}$ denotes the number of computed points on the solution curve, $\sum n_{\mathrm{it}}$ the total number of Newton-type iterations, $\overline{n}_{\mathrm{it}} = \frac{\sum n_{\mathrm{it}}}{n_{\mathrm{cont}}}$ the corresponding mean number of iterations per solution curve point, and $\sum n_{\mathrm{FC}}$ the total number of function
evaluations along the respective frequency-response curve.

\begin{table*}[ht]
    \centering
    \small
    \setlength{\tabcolsep}{4pt}
    \caption{Solver statistics for the adaptive continuation runs.
    The AFT-based formulation serves as reference.}
    \label{tab:continuation_statistics}
    \begin{tabular}{ll l rrrr}
        \toprule
        Test case & Parameters & Method
        & $n_{\mathrm{cont}}$
        & $\sum n_{\mathrm{it}}$
        & $\overline{n}_{\mathrm{it}}$
        & $\sum n_{\mathrm{FC}}$ \\
        \midrule

        \multirow{12}{*}{Cubic spring}
        & \multirow{2}{*}{$k_3=0.1,\;F_0=0.1$}
        & \shadedrow{gray!10}{AFT}{28}{54}{1.93}{82} \\
        & & \shadedrow{gray!10}{SFN}{28}{54}{1.93}{82} \\

        & \multirow{2}{*}{$k_3=0.1,\;F_0=0.2$}
        & \shadedrow{gray!20}{AFT}{44}{85}{1.93}{129} \\
        & & \shadedrow{gray!20}{SFN}{44}{85}{1.93}{129} \\

        & \multirow{2}{*}{$k_3=0.1,\;F_0=0.3$}
        & \shadedrow{gray!10}{AFT}{57}{106}{1.86}{163} \\
        & & \shadedrow{gray!10}{SFN}{57}{106}{1.86}{163} \\

        & \multirow{2}{*}{$k_3=-0.02,\;F_0=0.2$}
        & \shadedrow{gray!20}{AFT}{62}{78}{1.26}{140} \\
        & & \shadedrow{gray!20}{SFN}{62}{77}{1.24}{139} \\

        & \multirow{2}{*}{$k_3=0.01,\;F_0=0.2$}
        & \shadedrow{gray!10}{AFT}{52}{98}{1.88}{150} \\
        & & \shadedrow{gray!10}{SFN}{52}{98}{1.88}{150} \\

        & \multirow{2}{*}{$k_3=0.3,\;F_0=0.2$}
        & \shadedrow{gray!20}{AFT}{39}{75}{1.92}{114} \\
        & & \shadedrow{gray!20}{SFN}{39}{75}{1.92}{114} \\

        \midrule

        \multirow{12}{*}{Unilateral spring}
        & \multirow{2}{*}{$k_{\mathrm u}=10,\;g=1,\;F_0=0.1$}
        & \shadedrow{gray!10}{AFT}{76}{103}{1.36}{179} \\
        & & \shadedrow{gray!10}{SFN}{76}{90}{1.18}{166} \\

        & \multirow{2}{*}{$k_{\mathrm u}=1,\;g=1,\;F_0=0.1$}
        & \shadedrow{gray!20}{AFT}{131}{133}{1.02}{264} \\
        & & \shadedrow{gray!20}{SFN}{128}{131}{1.02}{259} \\

        & \multirow{2}{*}{$k_{\mathrm u}=100,\;g=0.5,\;F_0=0.1$}
        & \shadedrow{gray!10}{AFT}{44}{96}{2.18}{140} \\
        & & \shadedrow{gray!10}{SFN}{48}{112}{2.33}{160} \\

        & \multirow{2}{*}{$k_{\mathrm u}=100,\;g=2,\;F_0=0.1$}
        & \shadedrow{gray!20}{AFT}{109}{174}{1.60}{283} \\
        & & \shadedrow{gray!20}{SFN}{130}{282}{2.17}{412} \\

        & \multirow{2}{*}{$k_{\mathrm u}=100,\;g=1,\;F_0=0.1$}
        & \shadedrow{gray!10}{AFT}{69}{132}{1.91}{201} \\
        & & \shadedrow{gray!10}{SFN}{77}{161}{2.09}{238} \\

        & \multirow{2}{*}{$k_{\mathrm u}=100,\;g=1,\;F_0=0.05$}
        & \shadedrow{gray!20}{AFT}{57}{107}{1.88}{164} \\
        & & \shadedrow{gray!20}{SFN}{66}{150}{2.27}{216} \\

        \midrule

        \multirow{12}{*}{Jenkins element}
        & \multirow{2}{*}{$F_0=75,\;k_{\mathrm J}=10^7,\;\mu F_{\mathrm{N}}=106$}
        & \shadedrow{gray!10}{AFT}{100}{209}{2.09}{309} \\
        & & \shadedrow{gray!10}{SFN}{109}{221}{2.03}{330} \\

        & \multirow{2}{*}{$F_0=105,\;k_{\mathrm J}=10^7,\;\mu F_{\mathrm{N}}=106$}
        & \shadedrow{gray!20}{AFT}{107}{235}{2.20}{342} \\
        & & \shadedrow{gray!20}{SFN}{103}{247}{2.40}{350} \\

        & \multirow{2}{*}{$F_0=45,\;k_{\mathrm J}=10^7,\;\mu F_{\mathrm{N}}=53$}
        & \shadedrow{gray!10}{AFT}{100}{210}{2.10}{310} \\
        & & \shadedrow{gray!10}{SFN}{104}{222}{2.13}{326} \\

        & \multirow{2}{*}{$F_0=45,\;k_{\mathrm J}=10^7,\;\mu F_{\mathrm{N}}=530$}
        & \shadedrow{gray!20}{AFT}{106}{224}{2.11}{330} \\
        & & \shadedrow{gray!20}{SFN}{100}{224}{2.24}{324} \\

        & \multirow{2}{*}{$F_0=45,\;k_{\mathrm J}=10^7,\;\mu F_{\mathrm{N}}=106$}
        & \shadedrow{gray!10}{AFT}{103}{206}{2.00}{309} \\
        & & \shadedrow{gray!10}{SFN}{99}{205}{2.07}{304} \\

        & \multirow{2}{*}{$F_0=45,\;k_{\mathrm J}=5\times10^6,\;\mu F_{\mathrm{N}}=106$}
        & \shadedrow{gray!20}{AFT}{83}{186}{2.24}{269} \\
        & & \shadedrow{gray!20}{SFN}{78}{186}{2.38}{264} \\

        \bottomrule
    \end{tabular}
\end{table*}

\end{document}